\documentclass[a4paper]{article}
\usepackage{amsmath}
\usepackage{cite}
\usepackage{subfig}
\usepackage{soul}
\usepackage{color}
\usepackage{graphicx}
\usepackage{authblk}
\usepackage{float}
\providecommand{\keywords}[1]{\textbf{keywords: } #1}

\begin{document}
\title{Development of muon scattering tomography for a detection of reinforcement in concrete}

\author{Magdalena Dobrowolska$^{1}$\footnote{corresponding author: m.dobrowolska@bristol.ac.uk}, Jaap Velthuis$^{1,2,3}$, Anna Kopp$^1$, Chiara De Sio$^1$, Ruaridh Milne$^4$, Philip Pearson$^4$}

\affil{$^1$ School of Physics, HH~Wills Physics Laboratory, University of Bristol, Tyndall Avenue, BS8 1TL, Bristol, United Kingdom}
\affil{$^2$ School of Nuclear Science and Technology, University of South China,
No 28 West Changsheng Rd, Hengyang, China}
\affil{$^3$ Swansea University, Medical School, Swansea, SA2 8PP, United Kingdom}
\affil{$^4$ Cavendish Nuclear, Civil Structural \& Architectural (CS\&A) Department, Babcock Technology Centre (BTC), Unit 100A, Bristol Business Park, Stoke Gifford, BS16 1EJ, United Kingdom}

\maketitle

\begin{abstract}
Inspection of ageing, reinforced concrete structures is a world-wide challenge. Existing non-destructive evaluation
techniques in civil and structural engineering have limited penetration depth and don't allow to precisely ascertain the
configuration of reinforcement within large concrete objects. The big challenge for critical infrastructure (bridges, dams, dry docks, nuclear bioshields etc.) is understanding the internal condition of the concrete and steel, not just the location of the reinforcement. In most new constructions the location should be known and recorded in the as-built drawings, where these might not exist due to poor record keeping for older structures.

Muon scattering tomography is a non-destructive and non-invasive
technique which shows great promise for high-depth 3D concrete imaging. Previously, we have demonstrated that individual bars with a diameter of 33.7~$\pm$~7.3~mm at 50~cm depth can be located using muon scattering tomography. Here we present an improved method that exploits the periodicity of bar structures. With this new method, reinforcement with bars down to 6~mm thickness can be detected and imaged.

\end{abstract}

\keywords{3D imaging, concrete imaging, reinforcement location, bar location, muon scattering tomography, NDE technique}

\section{Introduction}
Old reinforced concrete structures may need to be inspected or replaced. Knowing the location of the steel is the first step towards determining the condition of the reinforcement. The key for assessing and substantiating the structure for life extensions is being able to state the design was built as planned (location and size of reinforcement) to a high quality (no voids from construction) and that the internal condition is satisfying (not degraded - reinforcement corrosion, cracking - beyond a critical value).  Current~non-destructive evaluation (NDE) scanning technology is based on the detection of reflected or transmitted electromagnetic, thermal or acoustic waves generated by a local source. The most used NDE techniques are magnetic imaging and ground penetrating radar (GPR) \cite{bungeySubsurfaceRadarTesting2004a, tarussovConditionAssessmentConcrete2013, rathodApplicabilityGPRRebar2019, changMeasurementRadiusReinforcing2009}, which can image bars with the diameters of 10-20~mm at depths of 100-500~mm. Low depth imaging, for depths at $<$~20~cm, can be performed with infrared thermographics 
\cite{milovanovicReviewActiveIR2016,maierhoferInfluenceConcreteProperties2007} and ultrasonics \cite{burrascanoUltrasonicNondestructiveEvaluation2014, lauretiDetectionRebarsConcrete2018}. These techniques are suitable for assessing element thickness and bar location, but precise estimation of bar size is still an unsolved problem \cite{tarussovConditionAssessmentConcrete2013}. Furthermore, a detailed testing below the first or
second reinforcement layer is often beyond the bounds of possibility because errors greatly increase with penetration depth and number of bars \cite{changMeasurementRadiusReinforcing2009}. 
Other NDE methods currently used are x-ray and neutron radiography \cite{duplessisReviewXrayComputed2019,zhangApplicationNeutronImaging2018}, which enable a high resolution and a high depth scanning. However, the use of active sources of radiation is a serious threat to human health.

Muon tomography is being investigated for many different challenges, relying on both Monte Carlo simulation studies and experiments. A traditional application of muon tomography is the characterization of nuclear waste drums and related security applications, where contents of concrete or bitumen filled waste drums are studied. Key issues here include the potential presence of gas bubbles in the matrix of the waste drum~\cite{dobrowolska2018novel} and identification of the material inside the 
drums~\cite{thomay2016passive,frazao2016discrimination,weekes2021material}. Security applications have been mainly focused on detection of lumps of high-Z material in cargo 
containers~\cite{thomay2013binned,4271541}, but work on the detection of explosives is ongoing as well~\cite{anbarjafari2021atmospheric}. 
Examples include experimental studies of imaging of concrete blocks~\cite{niederleithinger2020muon,checcia,durham}. 

Previously, we published a novel approach exploiting muon
scattering tomography (MST) to detect the presence and location of reinforcement bars \cite{dobrowolska2020towards}. This work has shown that a 100~cm long, singular bar with a diameter of 33.7~$\pm$~7.3~mm can be detected using three weeks of data taking at sea level. It was also shown that the signal has a monotonic dependence on the volume of the bar contained in the concrete drum. Moreover, the volume of the inclusion can be measured with a resolution of 5.4~$\pm$~0.3\%, and relative uncertainty below 10\%, for bar volumes above 2~500~cm$^{3}$. This outcome does not depend on the location of the bar. Bars as close as 6~cm apart can be detected as two individual items. However, the separation starts to be observable at a 4~cm distance. The approach also demonstrated to be suitable for imaging purposes, such as depicting bar structures in reinforced concrete. Differentiation between single and double layers of grids with bars diameters of 30~mm was also possible.

Here we present a new method that exploits the periodicity of bar structures, and is able to detect much smaller bar sizes within shorter time of data collection.\\

For most reinforced concrete structures, bars with diameters between 8~mm and 40~mm are used \cite{cannonsteelsltd}. The thinnest bars in use are 6~mm in diameter, whereas for
walls and bridges much thicker bars ($\ge$10~mm) are used.
The yield strength of the concrete depends strongly on the regular placement of the bars. Precise measurements are important for structural re-assessment to define a structural capacity or longerity of a concrete element or building structure. 
The spacing on most bar products is 10 or 20~cm \cite{reinforcementproductsonline}. Therefore, we have performed our studies with bars of a minimum diameter of 6~mm in a perfect grid of 7.5, 10, 15 and 20~cm.

\section{Muon scattering tomography (MST)}
Muon scattering tomography is a non-invasive method which shows great potential to generate high-depth 3D concrete images. MST uses cosmic rays as probes. Cosmic rays are high-energy charged particles which come to the Earth's atmosphere from outer space. In the atmosphere, cascades of new particles are produced. The main type of particles that reach sea level are muons.
Muons are identical to electrons, but 200 times heavier. Muons can go through large amounts of material as they do not scatter very much due to their high mass.
When traversing matter, Coulomb interactions take place between the muons and the nuclei of the material. As a result, muons undergo a series of scattering events and exit the material under an angle. 
The angular distribution of scattered muons can be
described by a Gaussian distribution with a mean of zero
and a standard deviation $\sigma_\theta$ described by \cite{eidelman2004review}:
\begin{equation}
\sigma _ {\theta} \approx \frac{13.6 \text{MeV}}{pc\beta} \sqrt{\frac{T}{X_{0}}} \  [1 + 0.038 \ln (\frac{T}{X_{0}})] 
\end{equation}
\begin{equation}
X_{0} \approx \frac{716.4A}{Z(Z+1) \ln (\frac{287}{\sqrt{Z}})} \  \  [\text{g} \cdot \text{cm} ^{-2}]
\end{equation}
where  \textsl{p} is muon's momentum; \textsl{$\beta$} is muon's speed divided by the speed of~light~\textsl{c}; \textsl{T} is the thickness of~the~material and $X_{0}$ its radiation length; \textsl{A} is the atomic weight of the medium in  g$\cdot$mol$^{-1}$. $\sigma_\theta$ depends on the atomic number $Z$ of the traversed material. 
Under the assumption that scattering occurs in a single location, and by reconstructing the incoming and 
outgoing trajectories of the muons, the scattering angle distribution can be reconstructed and thus 
information about the traversed material can be inferred. \\

\section{MST simulation}
In this work, we use Monte Carlo simulations of a MST system. The simulated MST system consists of detectors and a reinforced concrete block.
The muons were generated using the CRY library \cite{hagmann2007cosmic}, specifically developed for cosmic rays. GEANT4 \cite{agostinelli2003geant4} was used to simulate the propagation of the muons through detectors and scanned objects.

The simulated detector system consists of 3 sets of 2~$\times$~2~m$^2$ resistive plate chambers (RPCs) on one side of the volume under study and 3 sets of 2~$\times$~2~m$^2$ of RPCs on the other one. Three layers of detector sets on each side of the scanned object were chosen to provide a precise reconstruction of the muon direction and thus of the scattering angle. 

\begin{figure}[h]
	\centering
	\includegraphics[width=.79\columnwidth] {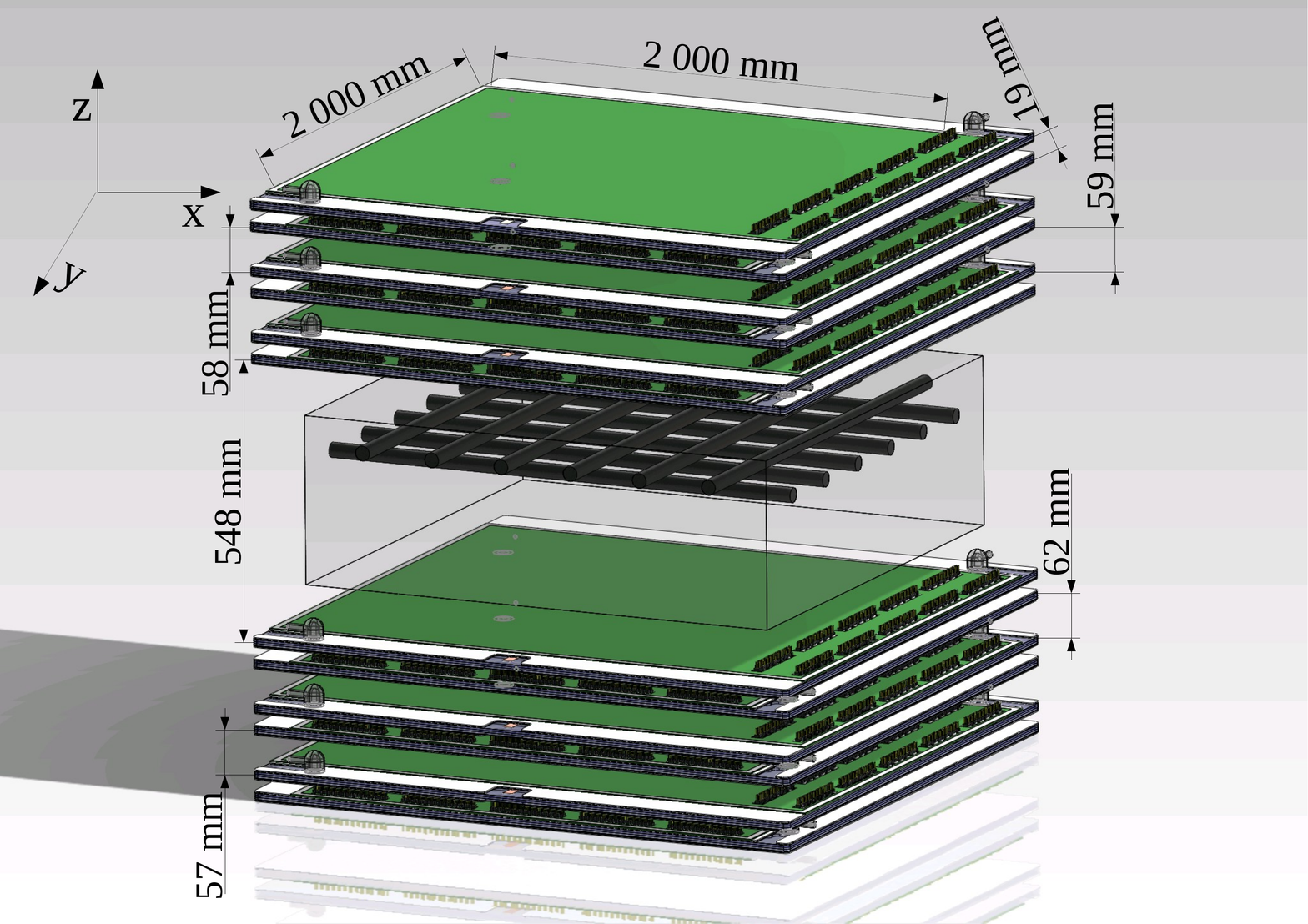}
	\caption{Schematic drawing of the simulated setup. RPCs are above and below the concrete block. Inside the concrete block a reinforcement grid is placed.}
	\label{simulation_sketch}
\end{figure}

RPCs consist of a gas volume enclosed between two glass panels over which a high voltage is applied. Muons, which are electrically charged, ionize the gas when 
traversing the detector. Due to the high voltage, a small avalanche is created locally. The measurable signal induced by the avalanche can be used to reconstruct the location where the muon traversed the RPC. The simulated RPCs performance was modeled using the performance of RPCs that were built for a container scanner prototype \cite{baesso2014toward, baesso2013high}. The RPCs had a pitch of 1.5~mm, which resulted in
a position resolution of approximately 450~$\mu$m. One pair of RPCs consists of both X and Y planes, orthogonal to each other, so that both x and y coordinates of the muon paths can be detected, and the muon paths calculated accurately. The thickness of one RPC is 6~mm. The X and Y planes are 19~mm apart and the distance between the RPC pairs is between 56 and 59~mm. The space between top and bottom detector layers, where an object can be scanned is 548~mm. 
The concrete block was placed in this volume. Reinforced concrete was modeled as a rectangular, concrete-filled object 
with dimensions of
200~cm~$\times$~200~cm~$\times$~34~cm. Inside the block, reinforcement bars were arranged in two orthogonal layers to form a single, reinforcement grid. Concrete was modeled as a material with a density of 2.3~g$\cdot$cm$^3$, and the steel reinforcement bars were simulated as iron bars with density of 7.87~g$\cdot$cm$^3$.
The density of concrete ranges from 2 to 2.5~g$\cdot$cm$^3$. A schematic drawing of the simulated setup is shown in figure~\ref{simulation_sketch}.

\begin{figure}[h]
	\centering
	\subfloat [Side (ZY) and front (ZX) views]{\label{fig: gsetup3ZY_ZX}
		\includegraphics[width=.49\columnwidth] {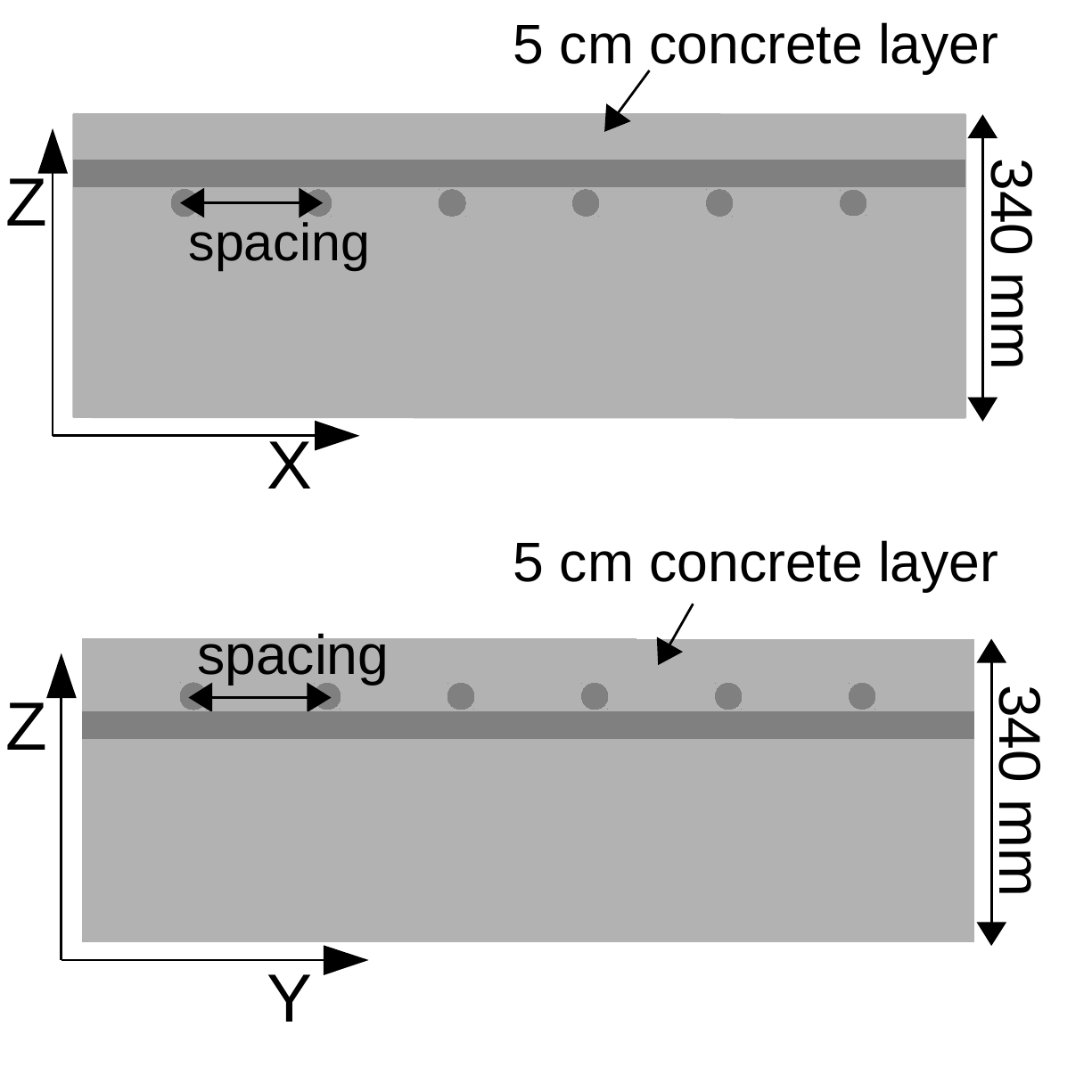}}
	\subfloat [Top (YX) view]{\label{fig: gsetup3YX}
		\includegraphics[width=.48\columnwidth] {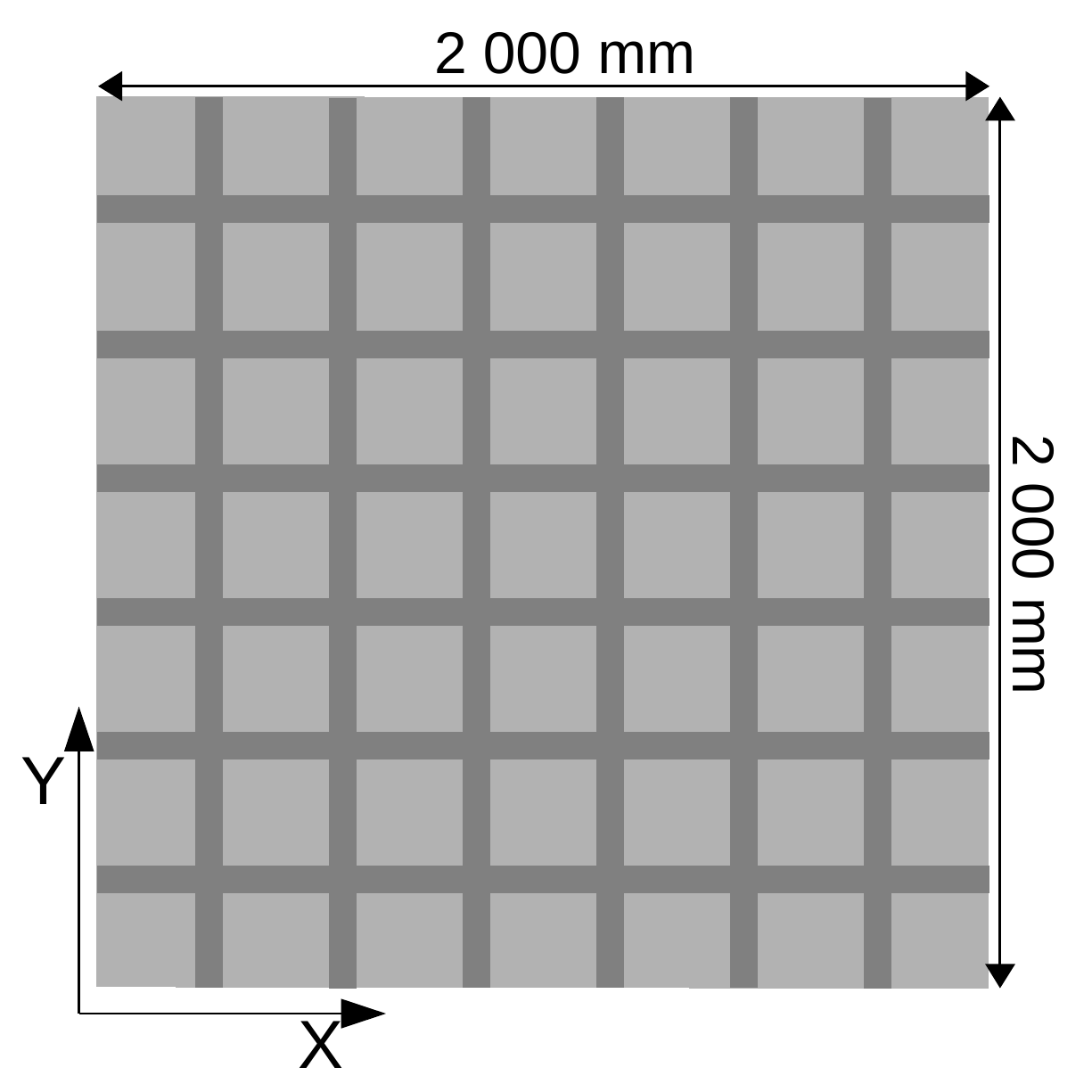}}
	\caption{A schematic drawing of the concrete block used in the simulations. Inside, two layers of reinforcement bars, forming a single grid were placed.}
	\label{scenarios}
\end{figure}

\label{simulation_scenarios}
Multiple scenarios were simulated to perform a detailed analysis. For all simulations the RPCs remained the same, whereas the contents of the concrete were changed. The diameter of the bars varied from 6~mm to 20~mm. Spacings of 7.5, 10, 15 or 20~cm were used. An example with a single layer of reinforcement grid is illustrated in figure~\ref{scenarios}. Figure~\ref{fig: gsetup3ZY_ZX} shows the ZX (front) and ZY (side) projection of the concrete block. The top (YX) projection is shown in figure~\ref{fig: gsetup3YX}.

\subsection{Bristol discriminator algorithm}
A number of MST imaging approaches have been developed. All of them use different 
ways to retrieve information from the incoming and outgoing muon tracks \cite{schultz2003cosmic, schultz2007statistical, stapleton2014angle}. We developed an algorithm which is described in detail in \cite{thomay2013binned}. It was successfully used to detect objects with a high atomic number (high-Z objects) in cargo containers \cite{thomay2013binned,thomay2015novel}, imaging of nuclear waste \cite{thomay2016passive}, as well as for
discrimination of high-Z materials in concrete-filled containers \cite{frazao2016discrimination} and detection of voids in concrete filled drums \cite{dobrowolska2018novel}. It was also demonstrated that bars with a diameter of 33.7~$\pm$~7.3~mm at 50~cm depth can be located using that approach \cite{dobrowolska2020towards}.

In our method incoming tracks are reconstructed using the three detector planes above and outgoing tracks using the three detector planes below the block. Subsequently, the hits are refitted under the assumption that the incoming and outgoing track meet in a vertex, where the scattering is assumed to have taken place in a single point. In reality this is not strictly true as the muon actually performs a random walk through the concrete block. However, the vertex assumption turns out to be a very useful one in practice. Our method relies on the "clusteredness" of high angle scatters in higher-Z materials: in high-Z materials the scattering angles tend to be larger, and larger scattering angles result in a well defined vertex. In addition, a higher number of large-angle scatters occur in higher-Z materials. This makes the method very sensitive to the detection of materials with a higher-Z inside an object of a lower-Z, or the other way around.

\begin{figure}[h]
	\centering
	\includegraphics[width=.49\columnwidth] {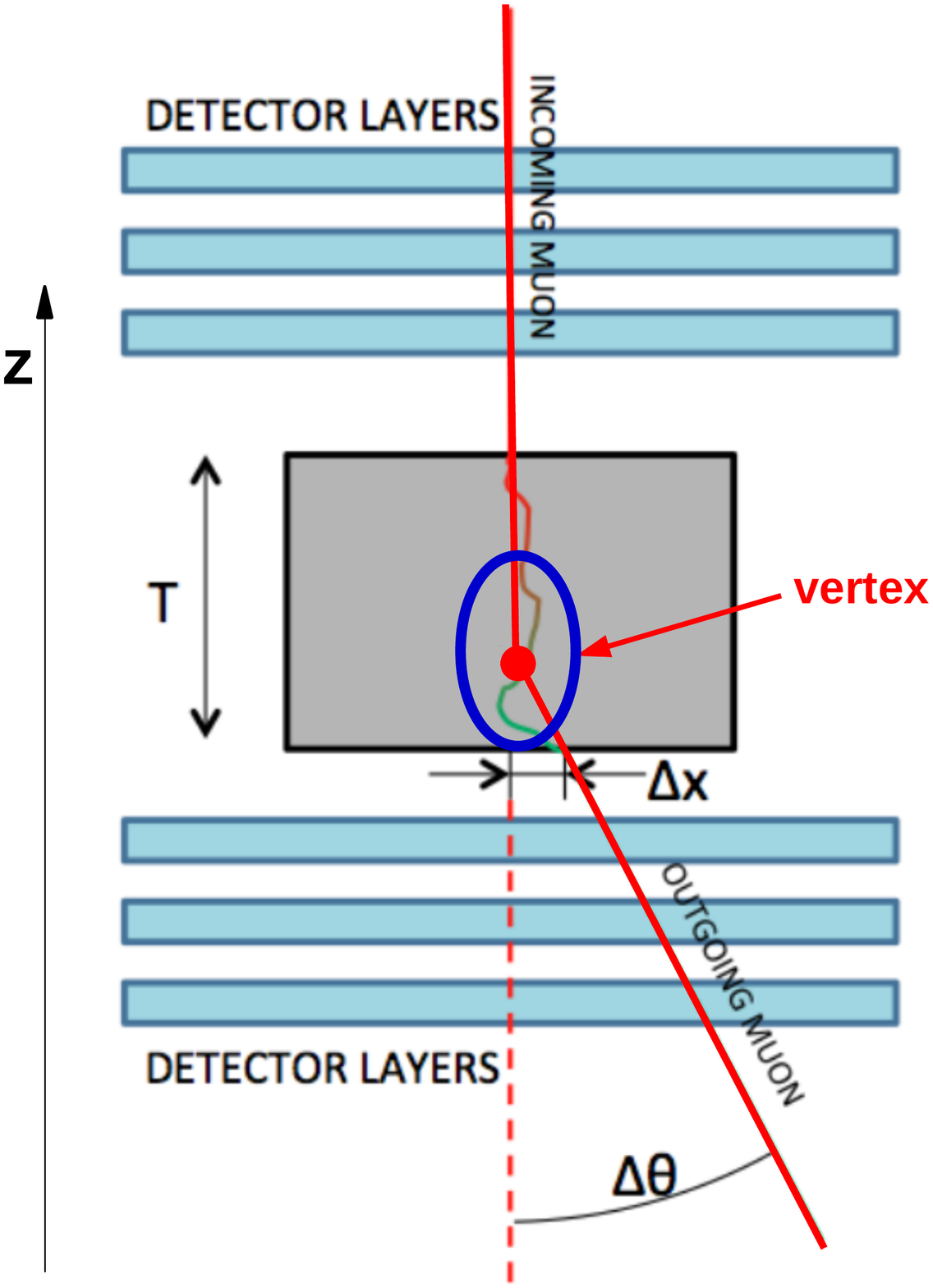}
	\caption{Incoming and outgoing muon tracks are reconstructed with RPCs. On their basis, the vertex is determined. }
	\label{muonScatteringPrinciple_with_vertex_final}
\end{figure}

\begin{figure}[H]
	\centering
	\includegraphics[width=.79\columnwidth] {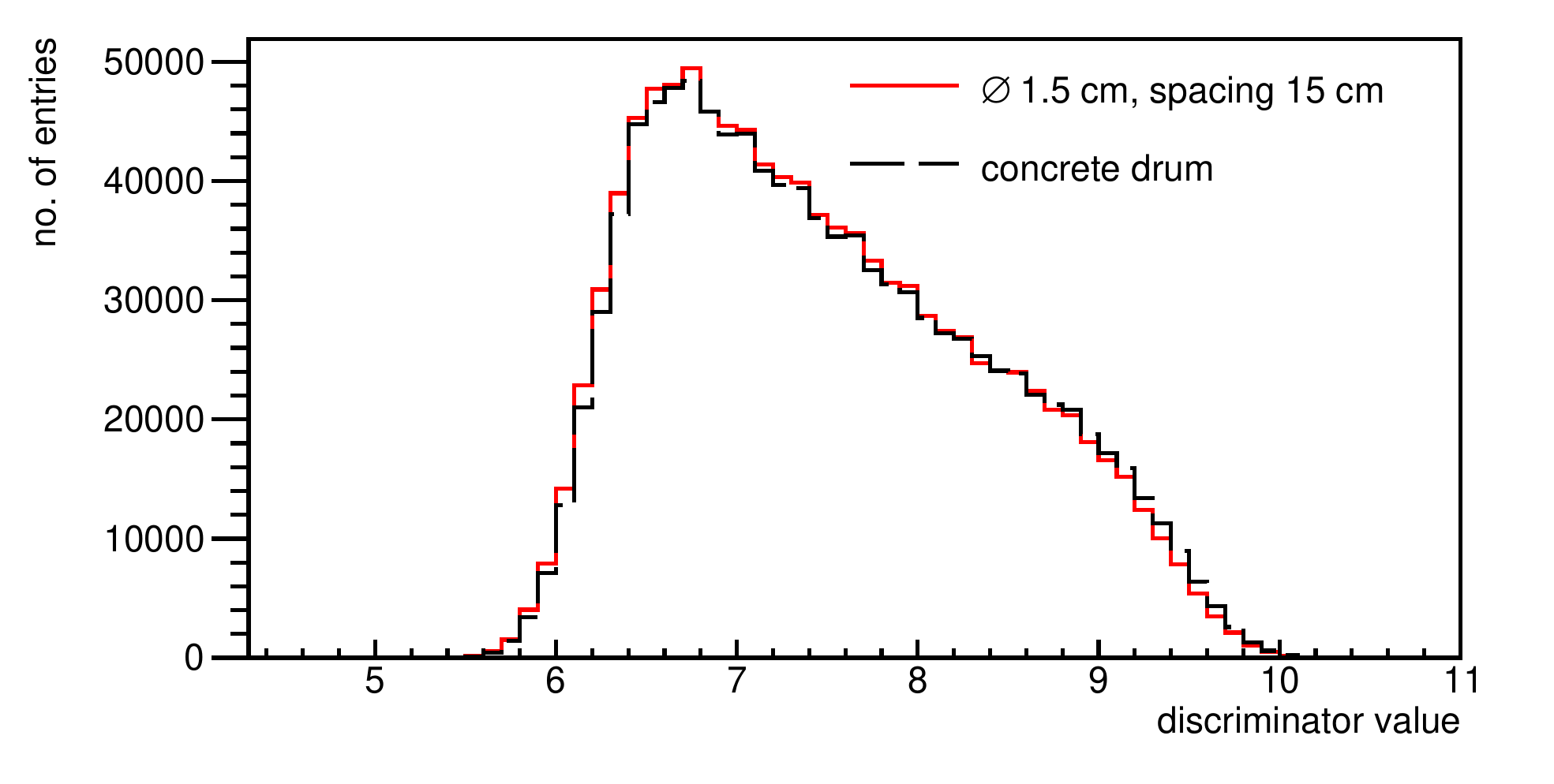}
	\caption{Discriminator distributions for a concrete block and a concrete block with single grid, where the bars' diameter was 1.5~cm.}
	\label{discrim_distrib}
\end{figure}

An example of  incoming, outgoing tracks and a vertex is shown in figure~\ref{muonScatteringPrinciple_with_vertex_final}. The scanned object 
is subdivided in voxels. A voxel size of 10~mm$\times$10~mm$\times$10~mm was used in this study. Each track-vertex-track combination is assigned to the voxel where the vertex is reconstructed. 
Since the method exploits the clusteredness of high angle scatters, only the $N$ most scattered tracks assigned to each voxel are considered in further analysis. $N$ of 30 was used for this analysis. For each pair of remaining vertices in that voxel, the weighted metric, $\widetilde{m_{ij}}$, is calculated:
\begin{equation}
\widetilde{m_{ij}} = \frac{\Vert \boldsymbol{V_{i} -V_{j}} \Vert}{\theta_{i} \cdot \theta_{j} }
\end{equation}
where \textsl{$\boldsymbol{V_{i}}$ }is the position of the vertex of muon \textsl{i}; $\theta_{i}$ is the corresponding scattering angle. The median of the weighted metric distribution is calculated for every voxel. The median of that distribution is then used as a discriminator \cite{thomay2013binned}. An example of the median discriminator distribution is shown in figure \ref{discrim_distrib}. In low atomic number (low-Z) materials, vertices are located at larger distances as high-angle scattering occurs less often than in denser materials. Hence, lower-Z materials correspond to higher discriminator values. In figure \ref{discrim_distrib} the discriminator distributions for a case with concrete only and a concrete block with reinforcement placed inside are shown. The reinforced block results in more low discriminator values.

\section{Detection of reinforcement bars}

\begin{figure} [h]	
	\centering
	\subfloat [single grid: front view]{\label{fig:ZX_1layer_1_5cm_907M_MT30}
		\includegraphics[width=.5\columnwidth] {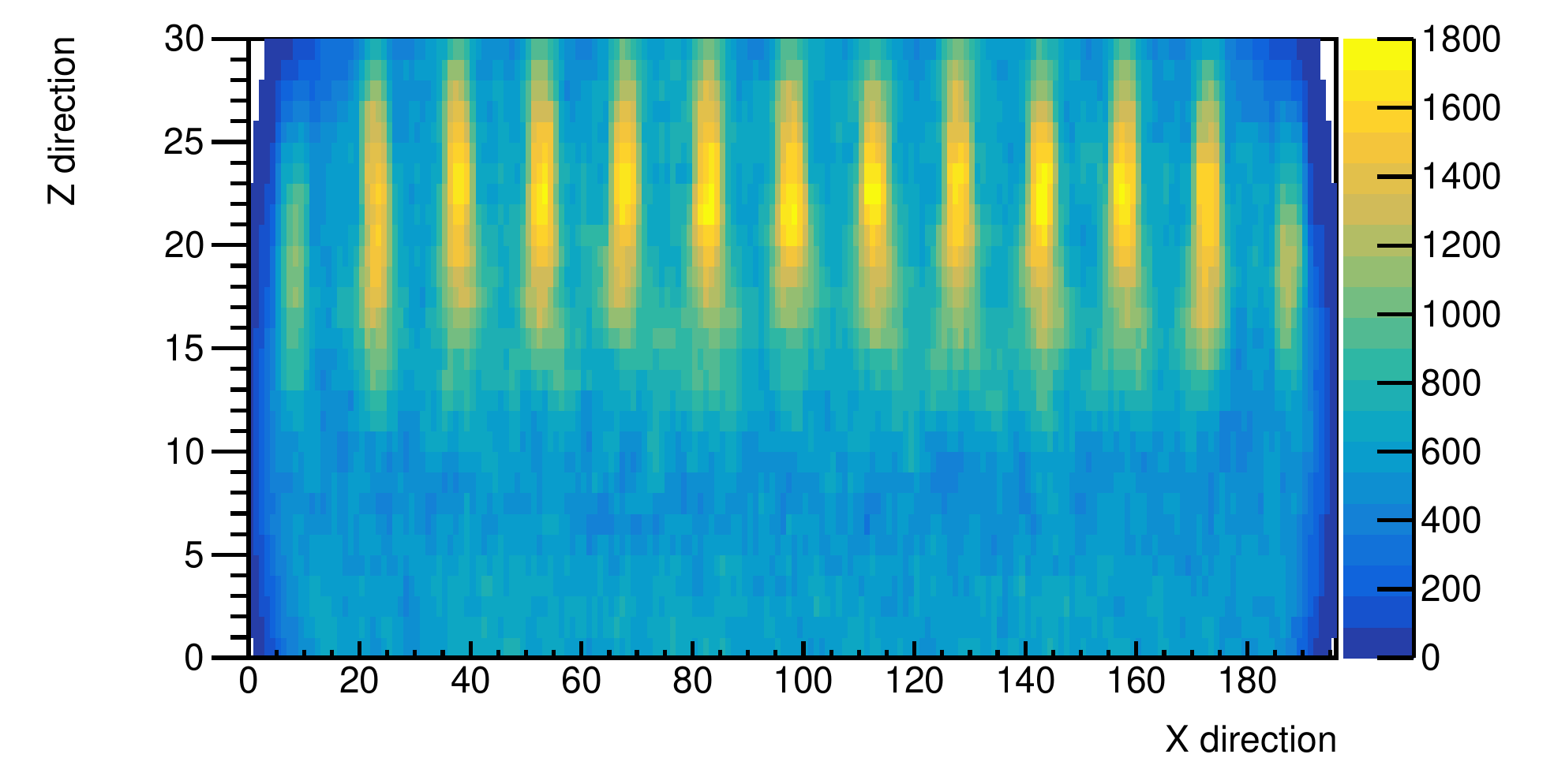}}	
	\centering
	\subfloat [single grid: front view]{\label{fig:ZX_1layer_1_5cm_907M_MT30_back_subtr}
		\includegraphics[width=.5\columnwidth] {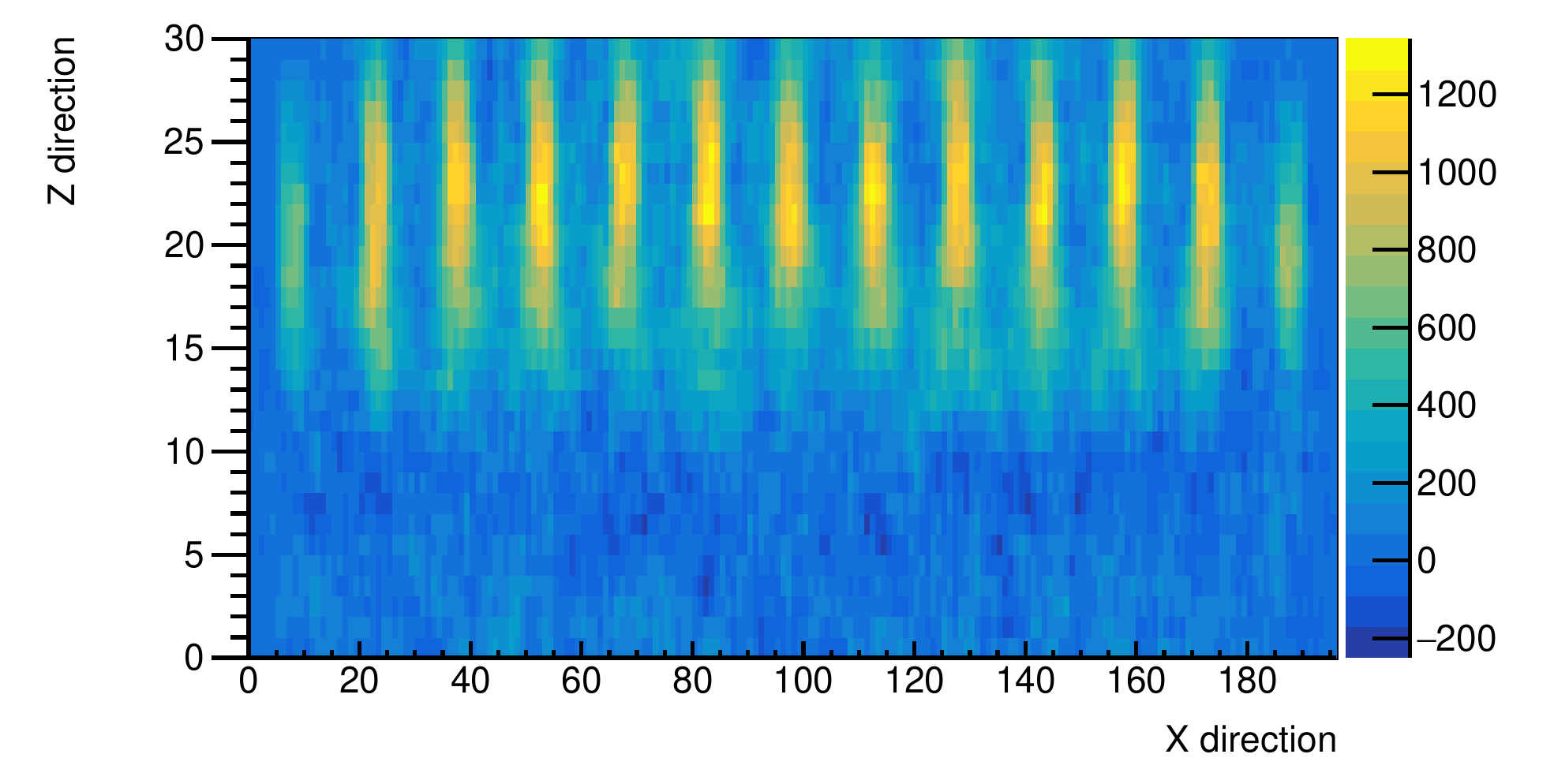}}
	\caption{Result of the reinforcement imaging before (left) and after background subtraction (right) for a single grid 
	using 1.5~cm diameter bars. }
	\label{fig: industrial_3_red}
\end{figure}

In this section a new detection algorithm will be presented which removes more background. An example of its performance will be shown using a concrete with a single grid made of bars with diameter of 1.5~cm, and a spacing of 15~cm. A background scenario was defined as pure concrete block. This method is an improved version of the algorithm published in~\cite{dobrowolska2020towards}. In the previous algorithm, for both background and reinforcement scenarios, for each voxel the discriminator is calculated. Next, a number of sub-volumes is created for each scenario and for each sub-volume, a discriminator distribution prepared. Then, for each bin of a discriminator distribution, the absolute difference between the
discriminator values of the block containing concrete and the scenario containing bars are taken. The resulting values are summed along the x, the y and the z-axis. The same study is repeated for every sub volume, resulting in three projections. A detailed description of the approach is in~\cite{dobrowolska2020towards}. An example of the front projection image (ZX) is shown in figure \ref{fig:ZX_1layer_1_5cm_907M_MT30}.

For the new method, all the above mentioned steps are done but the background subtraction was improved by generating two more concrete samples (the same size as bar-scenario ones) and performing 
the same analysis using the two concrete samples and thus generating final projection images for background only sample. Then, the background projections were subtracted, bin-by-bin, from the bar-concrete scenario projections. This method of background subtraction was chosen as the background is not linear and thus resulted in clear differences
between bar and concrete. An example of the projection image before and after additional background elimination is shown in figure \ref{fig: industrial_3_red}. Areas with higher signal clearly indicate the
presence of iron, while areas with lower signal show where the concrete is. Bars are clearly observable.

\subsection{Detection of the reinforcement grid}
\begin{figure}[H]
	\centering
	\includegraphics[width=.79\columnwidth] {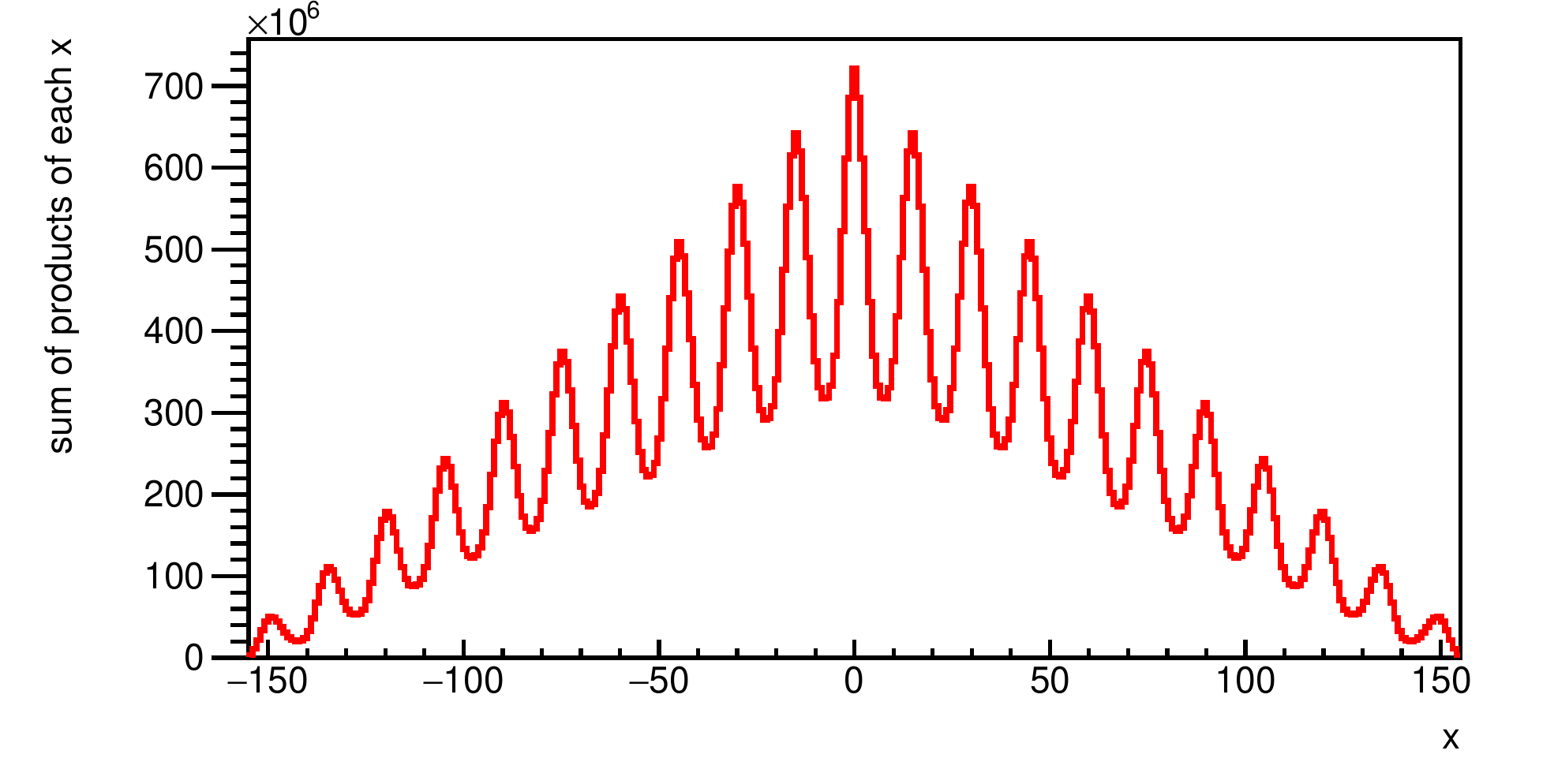}
	\caption{Result of the auto-correlation for the example case. }
	\label{CG_2layers_ZY_onemask}
\end{figure}

\begin{figure}[h]
	\centering
			\includegraphics[width=.48\columnwidth] {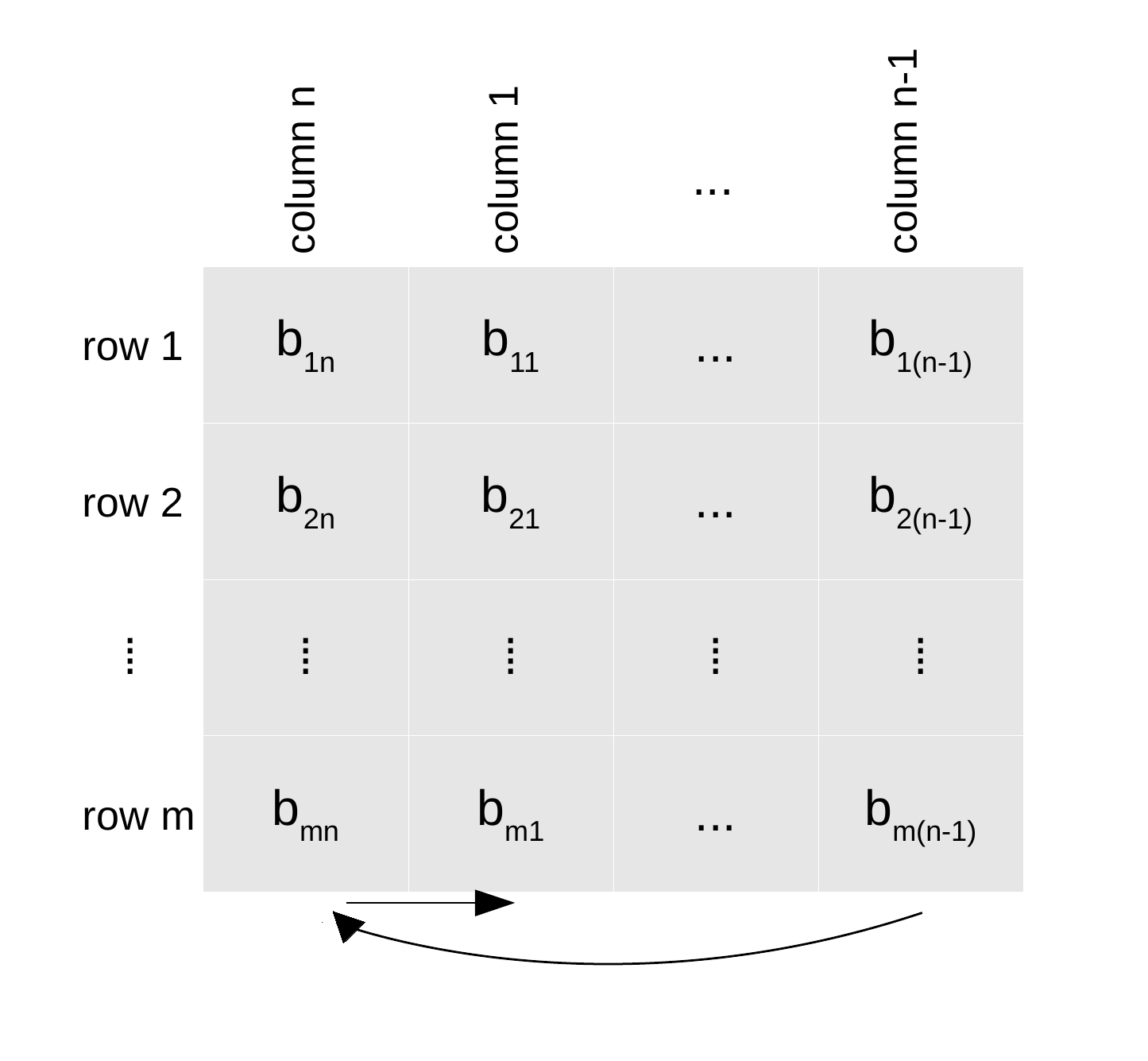}
	\caption{Illustration of the effect of the additional shift $\Delta_k$.}
	\label{fig:hist}
\end{figure}

\begin{figure}[h]
	\centering
	\includegraphics[width=.79\columnwidth] {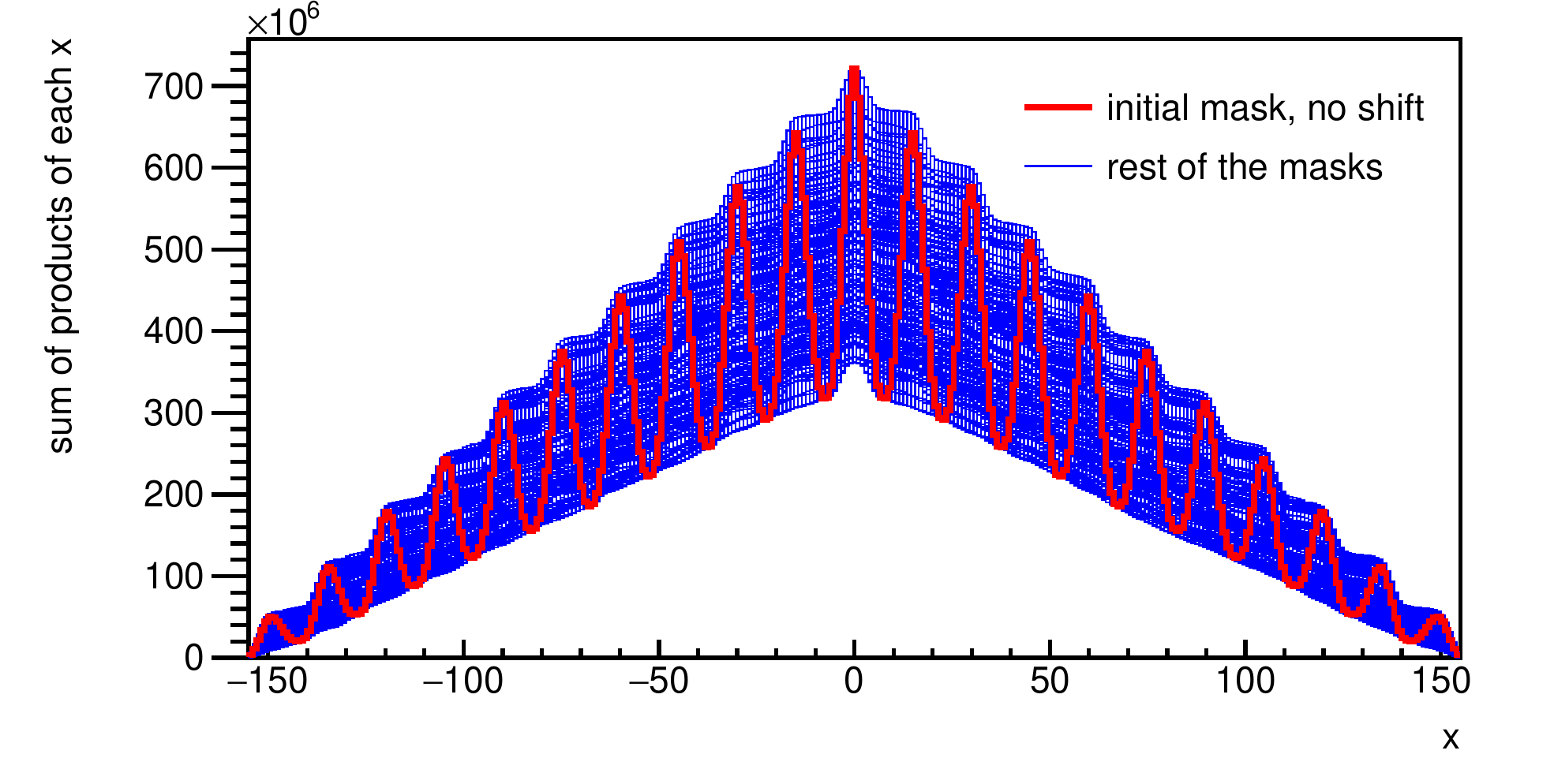}
	\caption{The auto-correlation for the signal sample and the complete series of auto-correlations
	for all values of $\Delta_k$.}
	\label{N2_1layer_1_5cm_side_view}
\end{figure}

\begin{figure}[h]
	\centering
		\includegraphics[width=.79\columnwidth] {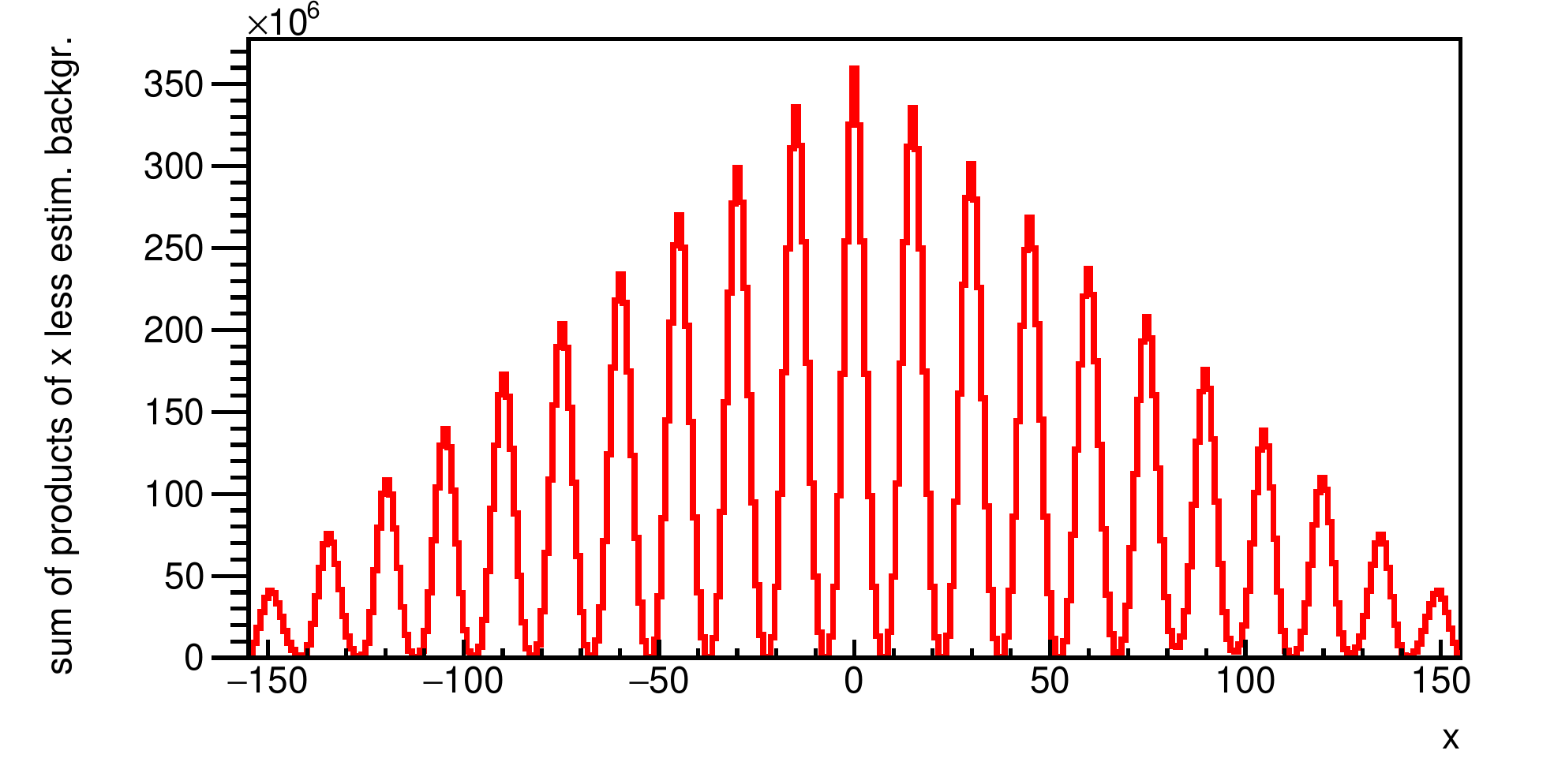}
	\caption{The auto-correlation for the signal sample after background subtraction.}
	\label{fig:after2backsub}
\end{figure}

From a structural engineering perspective, the reinforcement is always fixed in an orthogonal grid pattern. Due to its regular shape, the reinforcement provides a periodic signal. This periodicity can be exploited to lower the detection limit of the minimum diameter of the bars. This is done by calculating the auto-correlation of the 2D signal distributions (like figure~\ref{fig:ZX_1layer_1_5cm_907M_MT30_back_subtr}), and looking for peaks in the Fourier spectrum of 
the auto-correlation. 
The occurrence of peaks indicates the presence of a periodic structure and thus of bar. The peak locations yield the spacing, while the peak amplitude depend on the diameter of the bar. \\
In this method, a variation of the standard auto-correlation $R_s$ is used. Since there is periodicity in both the x and y direction, the auto-correlation is only evaluated in one direction, i.e. the x direction, as:
\begin{equation}
R_s(\tau)=\int \limits_{y_{min}}^{y_{max}} \int \limits_{x_{min}}^{x_{max}} f(x',y')f(x'+\tau,y')dx'dy'
\label{eq:autcorr}
\end{equation}
The calculation was limited to the area within the acceptance of the detector, and the presence of the reinforcement
 i.e., {X$ \in<$20;175$>$, Y$\in<$10;29$>$} in figure~\ref{fig:ZX_1layer_1_5cm_907M_MT30_back_subtr}.
\label{amplification}
The result of the auto-correlation for the example case is shown in figure \ref{CG_2layers_ZY_onemask}.  The triangular shape is due to the variation of the overlapping area. 
It is observed as a triangular background and the triangular dependence of the amplitude of the periodic structure.
The periodic structure is due to the reinforcement spacing.
Before the Fourier transformation, that triangular background needs to be subtracted. To estimate it, the complete series of auto-correlation, $R_{b,k}$ is calculated, where:
\begin{equation}
R_{b,k}=\int \limits_{y_{min}}^{y_{max}} \int \limits_{x_{min}}^{x_{max}} f(x',y')f(x'+\Delta_k+\tau,y')dx' dy'
\label{eq:autcorr_bg}
\end{equation}
Here the function is shifted by an additional $\Delta_k$, where $k$ indicates the number of pixels the image has been shifted. The shift occurs in a rolling mode, i.e. when a column is shifted further than $x_{max}$, it is placed in column 0. This is illustrated in figure~\ref{fig:hist}. 
This procedure leads to a shift in the peaks, but the underlying triangular background shape remains the same, as can be seen in  
figure~\ref{N2_1layer_1_5cm_side_view}. 
For each bin in the signal auto-correlation, the minimum value of $R_{b,k}$ is subtracted.   
The result is shown in figure~\ref{fig:after2backsub}. The graph still displays the triangular pattern in the 
amplitude, but the triangular background under the function is removed.

\begin{figure}[h]
	\centering
		\includegraphics[width=.95\columnwidth] {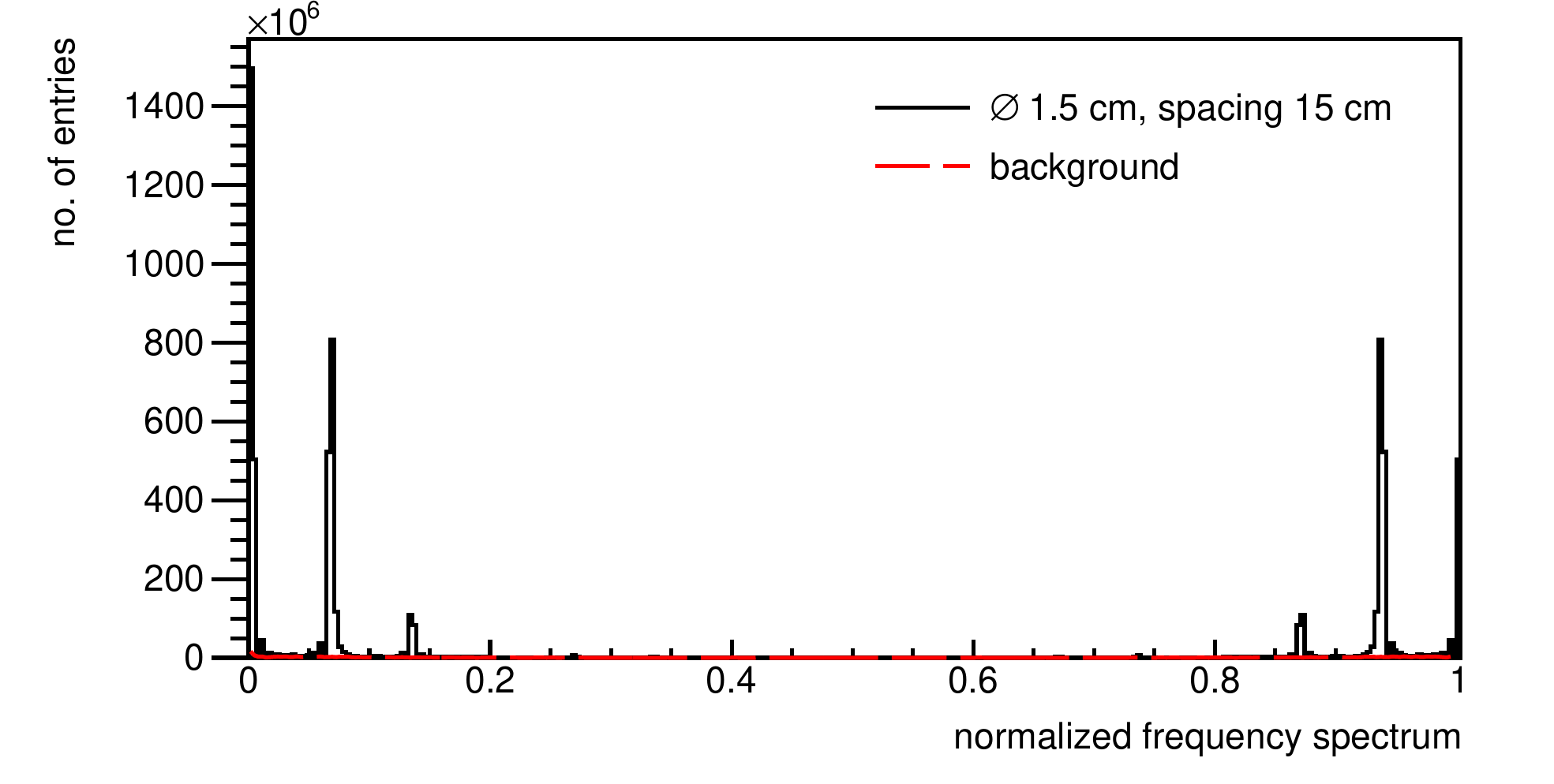}
	\caption{Fourier spectrum after background subtraction for the example case with 1.5~cm diameter bars and a 15~cm spacing.}
	\label{fig: frequency}
\end{figure}

Next, the background-subtracted signal is Fourier transformed using an interface class for Fast Fourier Transforms, TVirtualFFT, provided by ROOT, a data analysis package developed at CERN \cite{FFT}. Figure~\ref{fig: frequency} shows the result of the FFT. The spectrum shows a series of clearly visible distinct
peaks. Their position is determined by the bar spacing and the peak amplitudes are related to the 
bar diameter. A background scenario, consisting of a concrete block without reinforcement, does not display any peaks.
These results show that the method is capable of detecting bar  grids.\\
 
\subsection{Variation of the bar diameter}
\begin{figure}[h]
	\centering
	\includegraphics[width=.95\columnwidth] {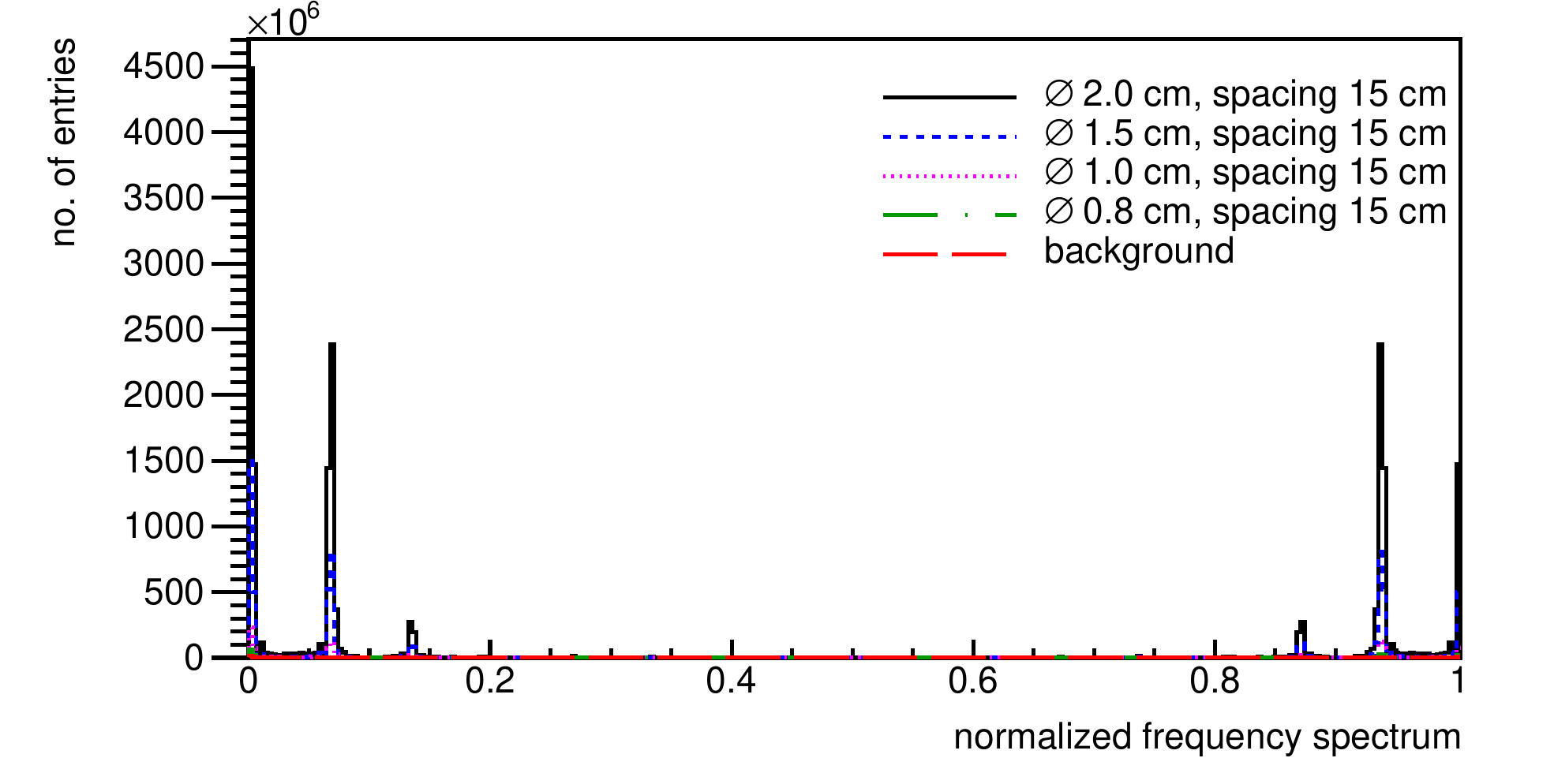}
	\caption{Normalized frequency spectrum for a reinforcement grid made of 8, 10, 15 or 20~mm diameter bars and spacing of 15~cm.}
	\label{fig: N2_1layer_mesh15_front_view_402M_30MT_frequency_1_5_1_0_0_8}
\end{figure}

\begin{figure}[h]
	\centering
	\includegraphics[width=.95\columnwidth] {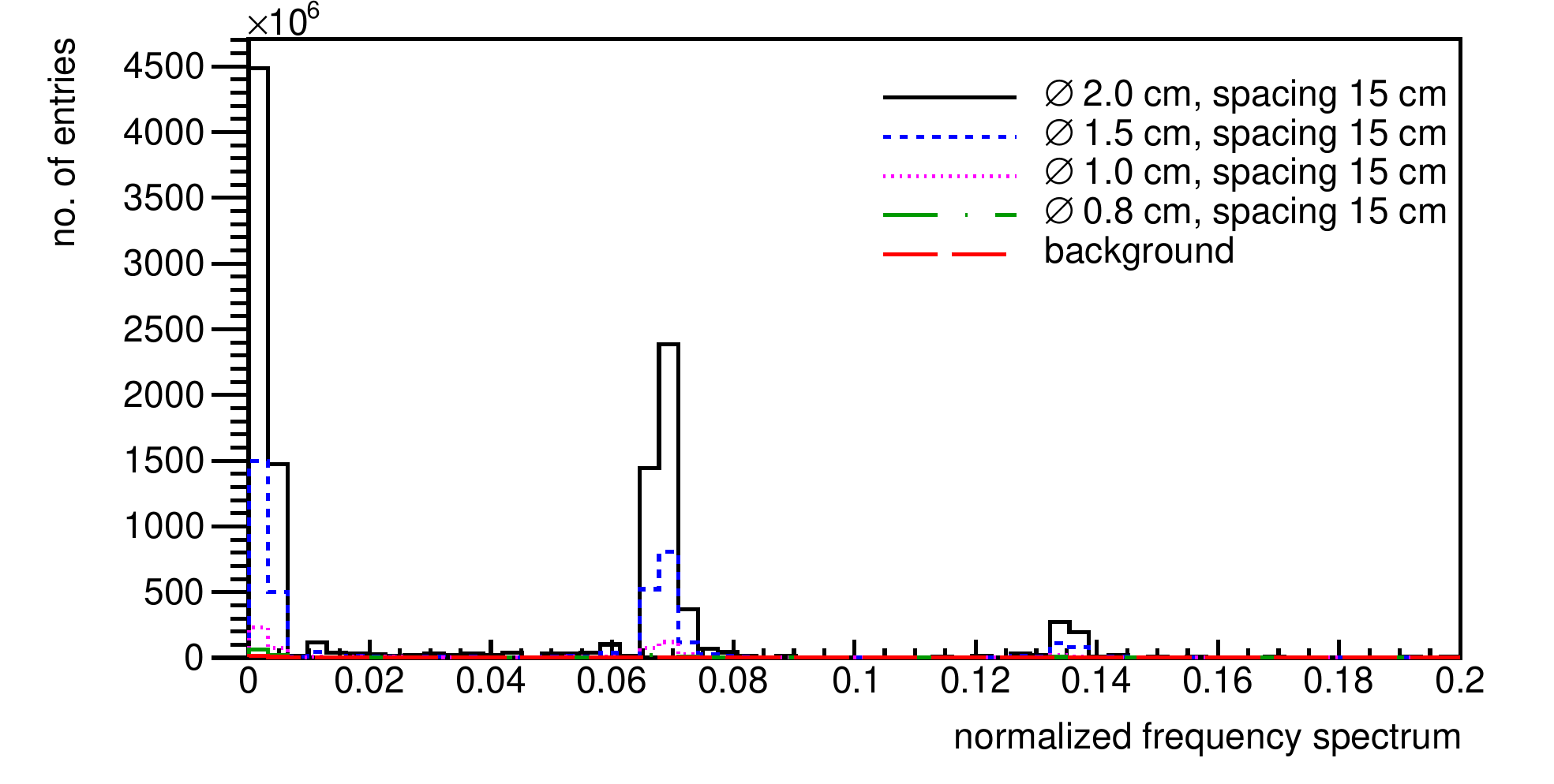}
	\caption{Zoomed, normalized frequency spectrum for a reinforcement grid made of 8, 10, 15 or 20~mm diameter bars and spacing of 15~cm.}
	\label{fig: N2_1layer_mesh15_front_view_402M_30MT_frequency_1_5_1_0_0_8_zoomed}
\end{figure}

\begin{figure}[h]
	\centering
	\includegraphics[width=.95\columnwidth] {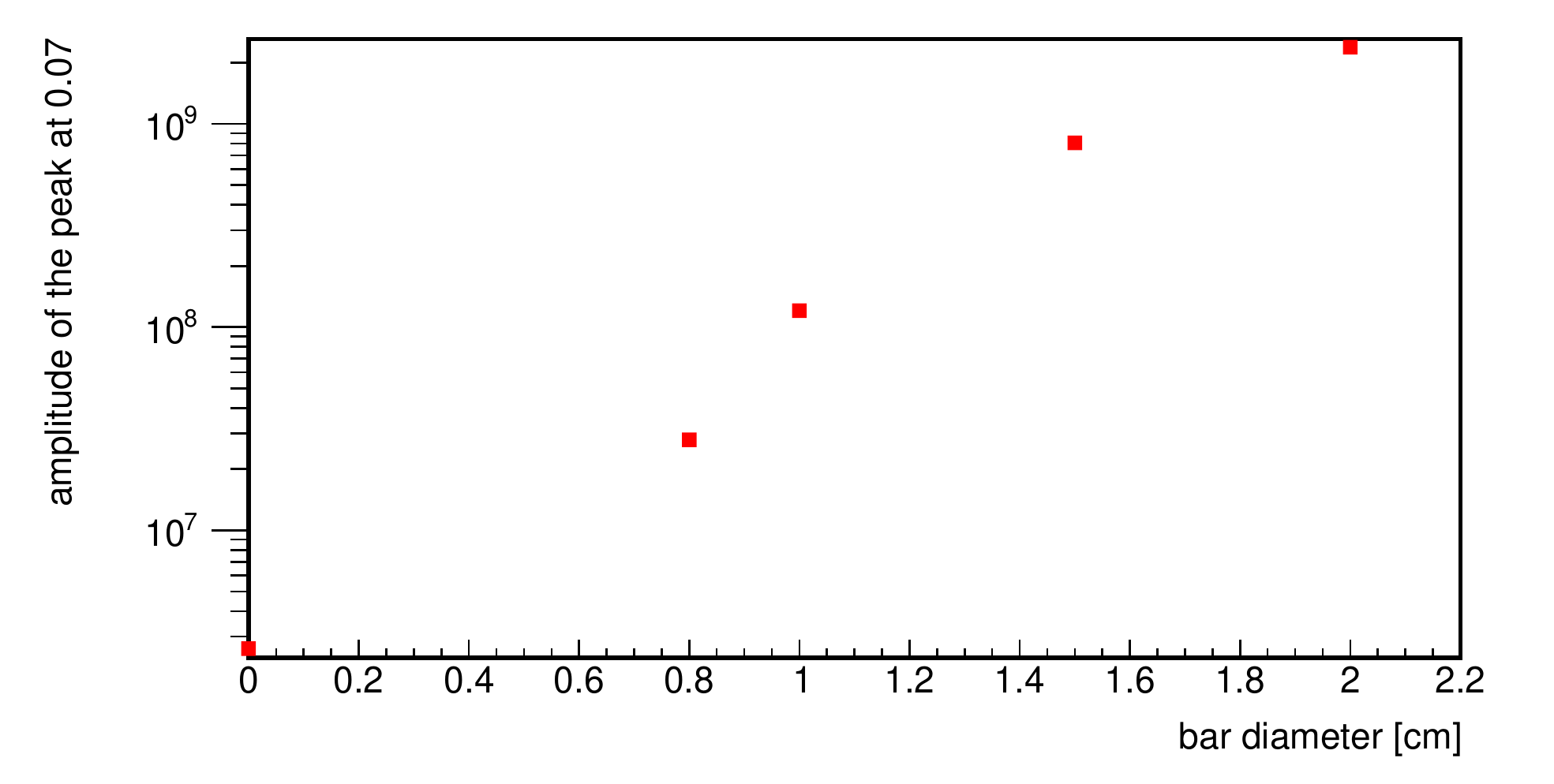}
	\caption{Amplitude of the peak at 0.07 of the normalized frequency distribution for different size of the bar diameter and spacing fixed at 15 cm. There is a clear dependence between the bar diameter and the amplitude; the bigger the diameter the higher the amplitude. }
	\label{fig:amplitude}
\end{figure}

The peak locations of the Fourier transform depend on the spacing, while the peak amplitude
is related to the bar diameter. 
Figure \ref{fig: N2_1layer_mesh15_front_view_402M_30MT_frequency_1_5_1_0_0_8} shows
the Fourier transforms for reinforcement grid made of 20, 15, 10 and 8~mm diameter bars with a 15~cm spacing, see figure \ref{fig: N2_1layer_mesh15_front_view_402M_30MT_frequency_1_5_1_0_0_8_zoomed} for a zoomed version of the plot. As expected, having the same spacing, the peaks occur always at the same normalized frequency values. With decreasing bar diameter, the amplitude of the peaks also decreases. This is presented more clearly in figure~\ref{fig:amplitude}, which shows amplitude of the peak at 0.07 of the normalized frequency plot as a function of bar diameter. The amplitude strongly increases with increasing the diameter. Hence, the bar diameter can be measured based on normalized frequency spectrum.


\subsection{Variation of the spacing}
\begin{figure}[H]
	\centering
	\includegraphics[width=.95\columnwidth] {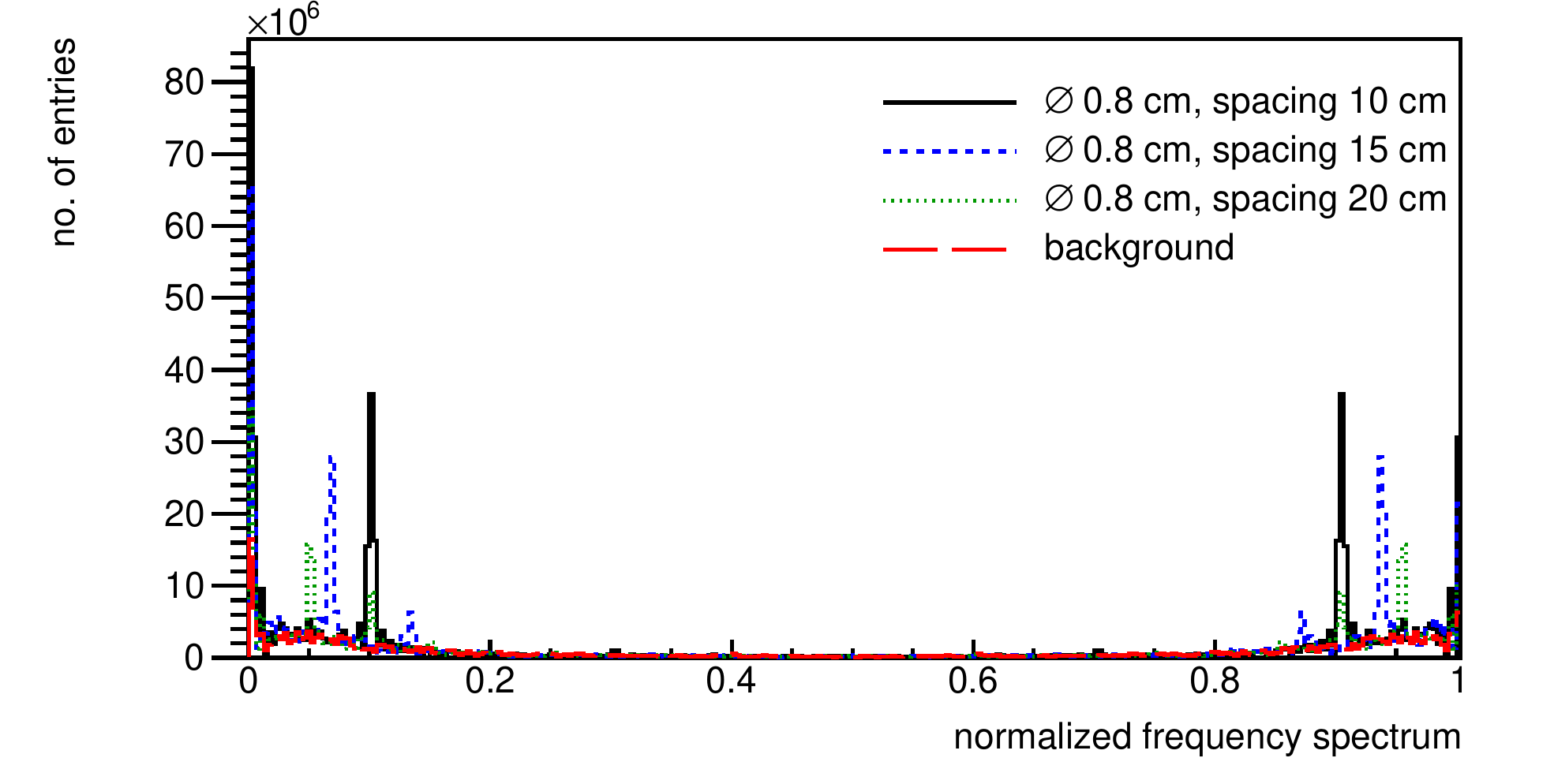}
	\caption{Normalized frequency spectrum for a reinforcement made of 8~mm diameter bars, and spacing of 10~cm, 15~cm or 20~cm, respectively.}
	\label{fig: N2_1layer_0_8cm_front_view_402M_30MT_frequency_diffMesh}
\end{figure}

\begin{figure}[H]
	\centering
	\includegraphics[width=.95\columnwidth] {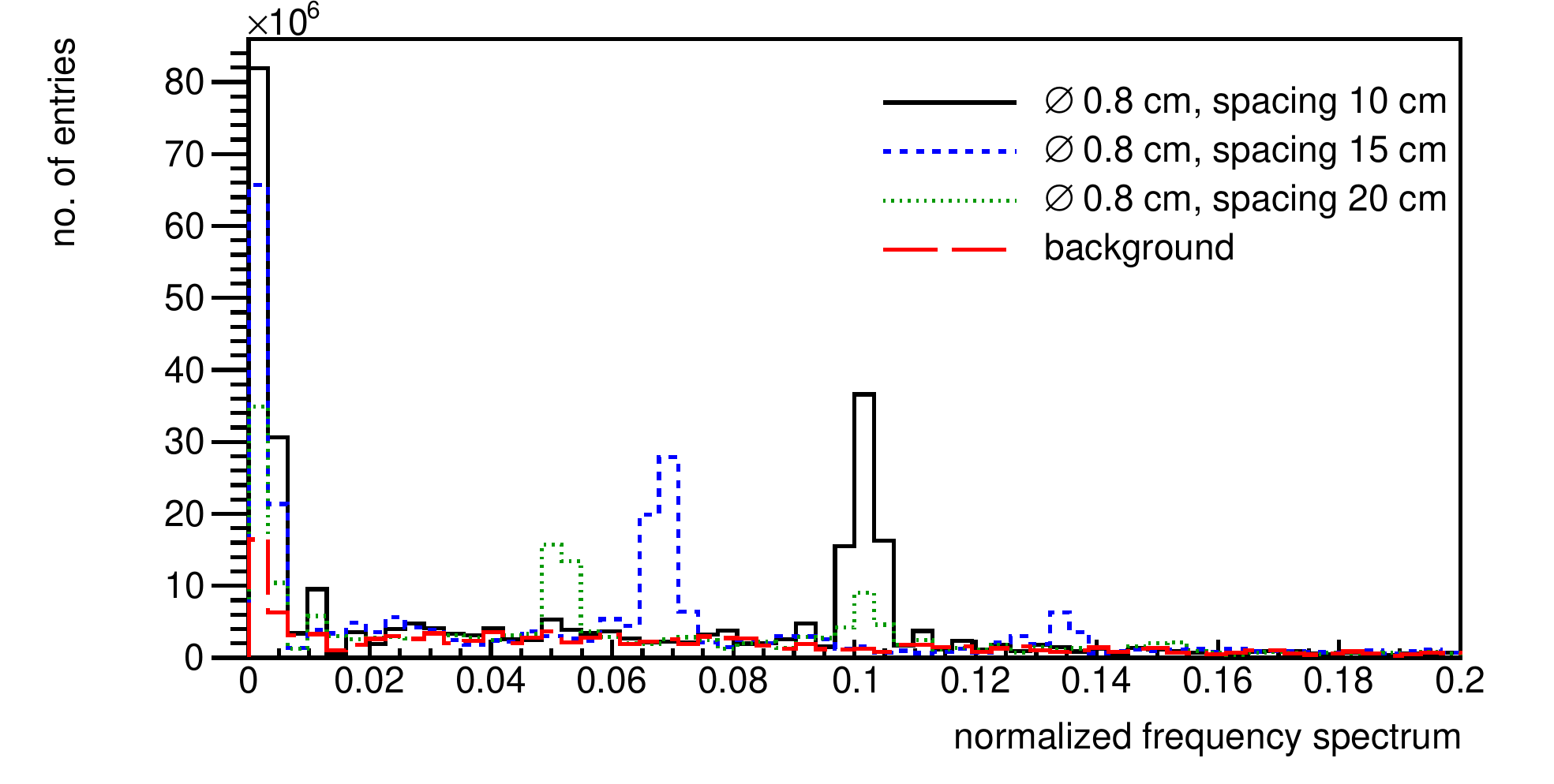}
	\caption{Zoomed, normalized frequency spectrum for a reinforcement made of 8~mm diameter bars, and spacing of 10~cm, 15~cm or 20~cm, respectively.}
	\label{fig: N2_1layer_0_8cm_front_view_402M_30MT_frequency_diffMesh_zoomed}
\end{figure}

In the previous results, a spacing of 15 cm was used. Figure \ref{fig: N2_1layer_0_8cm_front_view_402M_30MT_frequency_diffMesh} shows the Fourier spectrum for the reinforcement using 8~mm diameter bars with spacing of 10, 15 and 20~cm, see figure \ref{fig: N2_1layer_0_8cm_front_view_402M_30MT_frequency_diffMesh_zoomed} for a zoomed version of that figure. All of the cases are clearly distinguishable from the background scenario. Moreover, the peaks are located at different frequencies, which makes this method suitable for the estimation of the spacing.		

\subsection{Limits of the method}
\begin{figure}[H]
	\centering
	\includegraphics[width=.95\columnwidth] {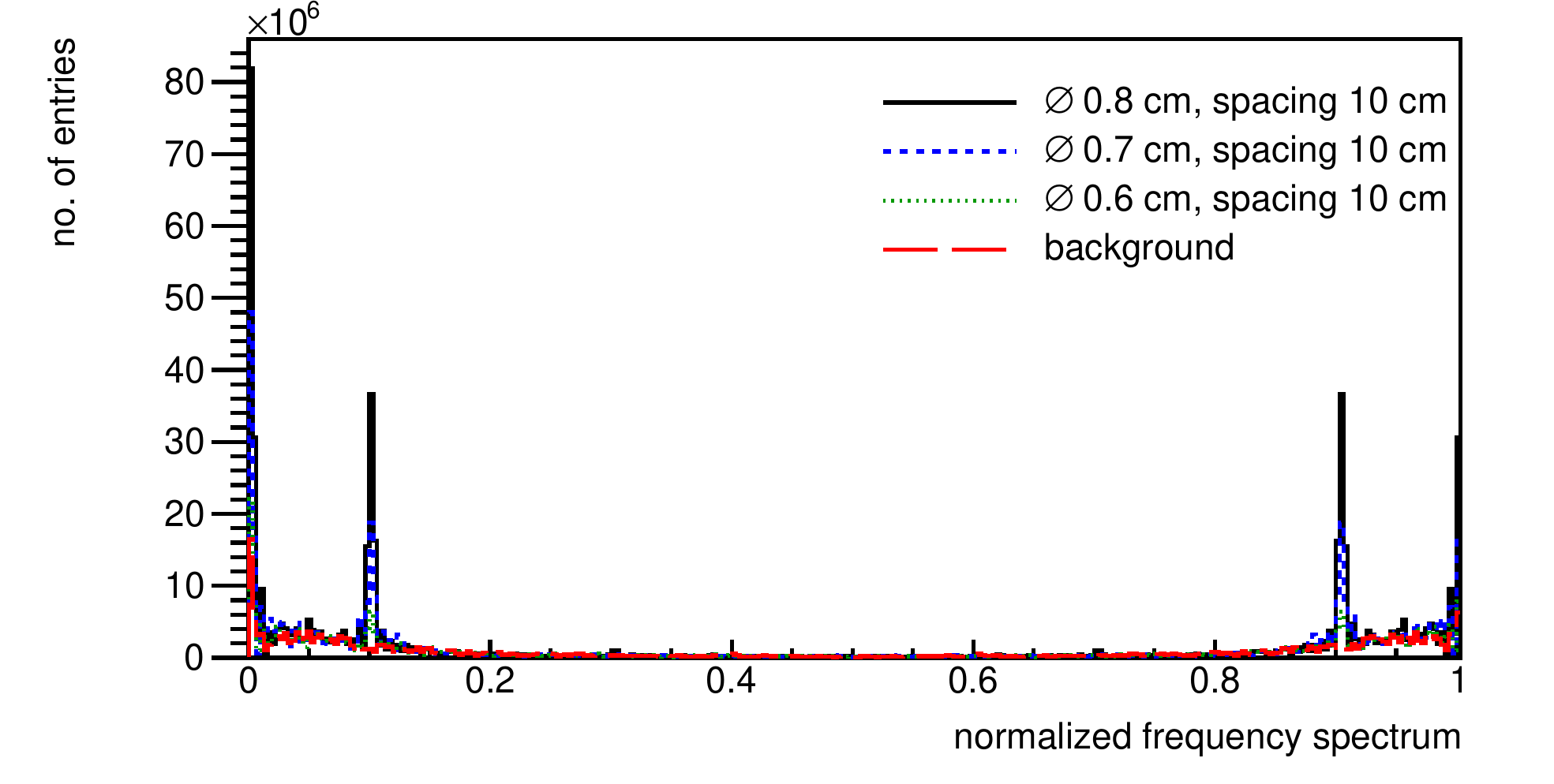}
	\caption{Normalized frequency spectrum for a reinforcement made of 6, 7 or 8~mm diameter bars and spacing of 10~cm.}
	\label{fig: N2_1layer_mesh10_front_view_402M_30MT_frequency_0_6_0_7_0_8}
\end{figure}

\begin{figure}[H]
	\centering
	\includegraphics[width=.95\columnwidth] {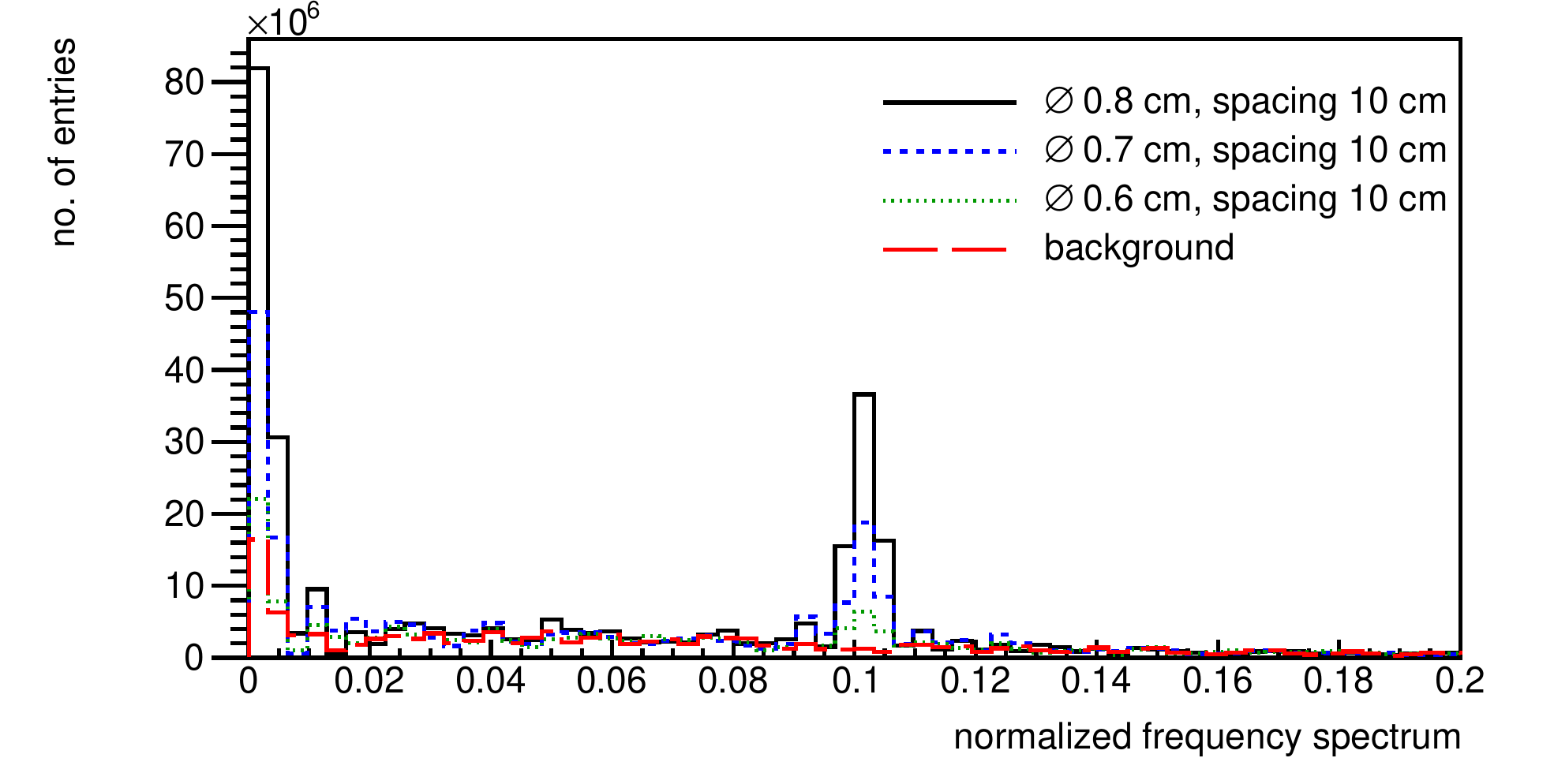}
	\caption{Zoomed, normalized frequency spectrum for a reinforcement made of 6, 7 or 8~mm diameter bars and spacing of 10~cm.}
	\label{fig: N2_1layer_mesh10_front_view_402M_30MT_frequency_0_6_0_7_0_8_zoomed}
\end{figure}

\begin{figure}[H]
	\centering
	\includegraphics[width=.95\columnwidth] {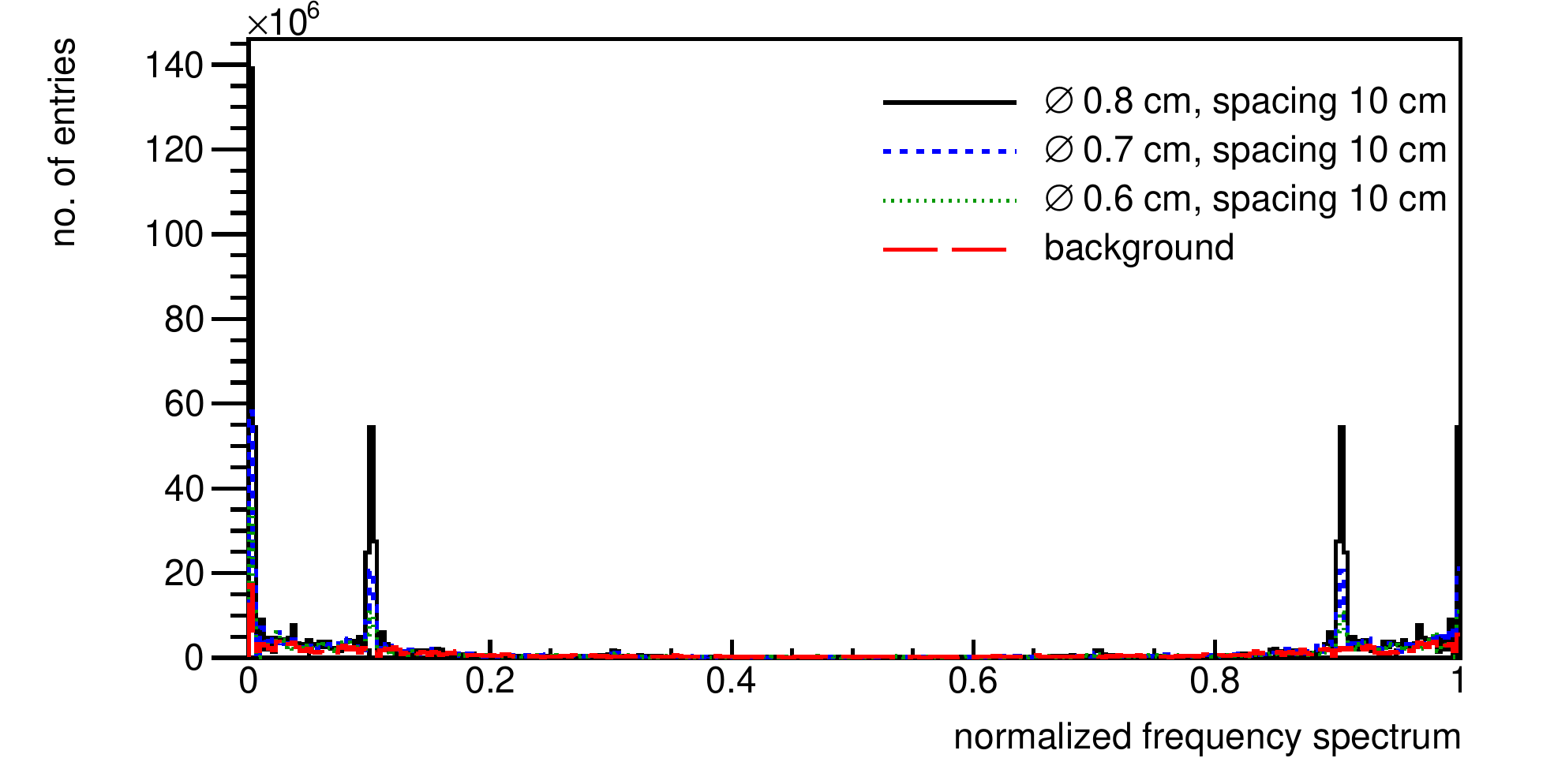}
	\caption{Normalized frequency spectrum for a reinforcement made of 6, 7 or 8~mm diameter bars and spacing of 10~cm. Time of data taking was increased to two weeks.}
	\label{fig: N2_1layer_mesh10_front_view_804M_30MT_frequency_0_6_0_7_0_8}
\end{figure}

\begin{figure}[H]
	\centering
	\includegraphics[width=.95\columnwidth] {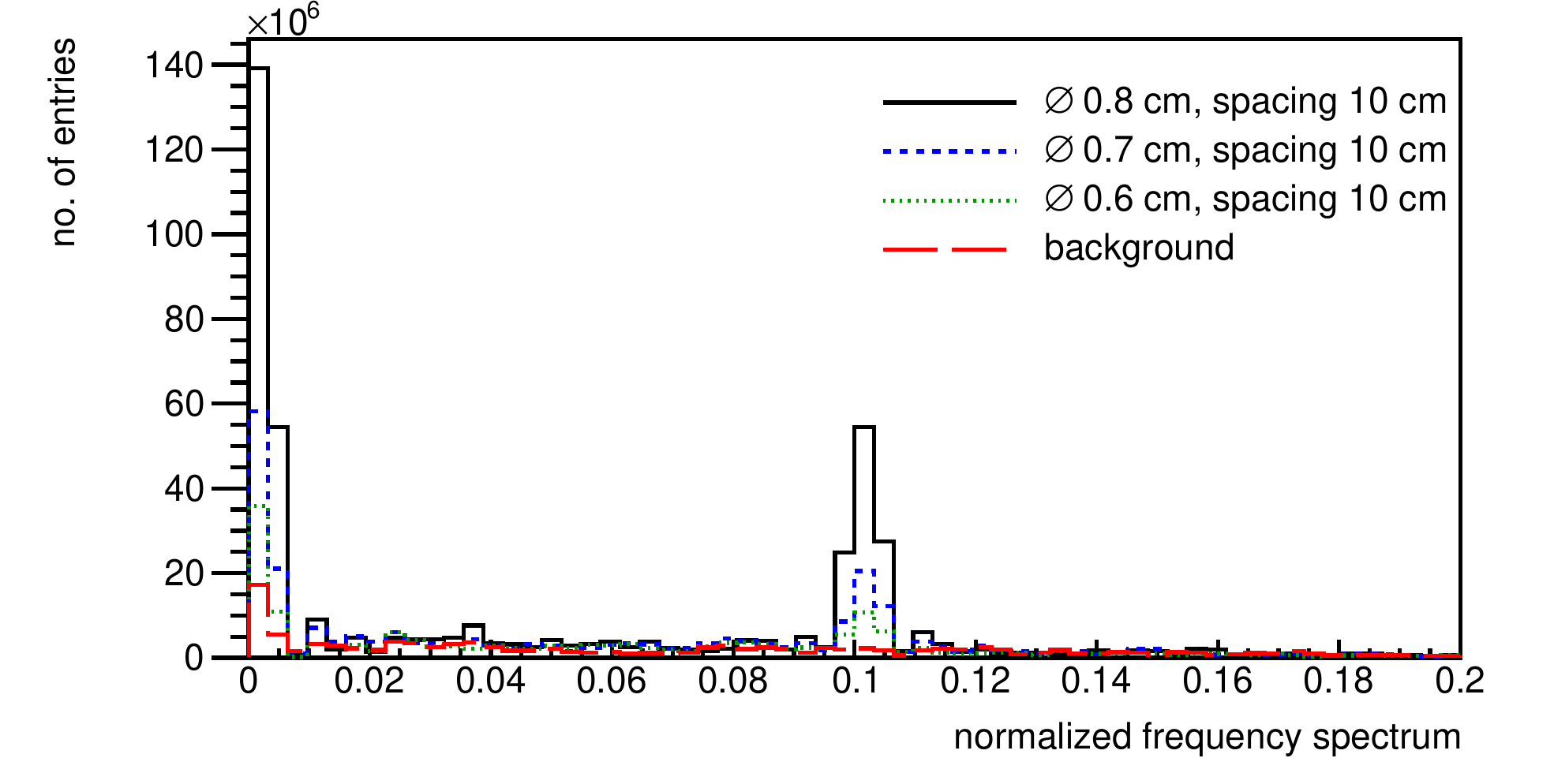}
	\caption{Zoomed, normalized frequency spectrum for a reinforcement made of 6, 7 or 8~mm diameter bars and spacing of 10~cm. Time of data taking was increased to two weeks.}
	\label{fig: N2_1layer_mesh10_front_view_804M_30MT_frequency_0_6_0_7_0_8_zoomed}
\end{figure}

\begin{figure}[h]
	\centering
	\includegraphics[width=.95\columnwidth] {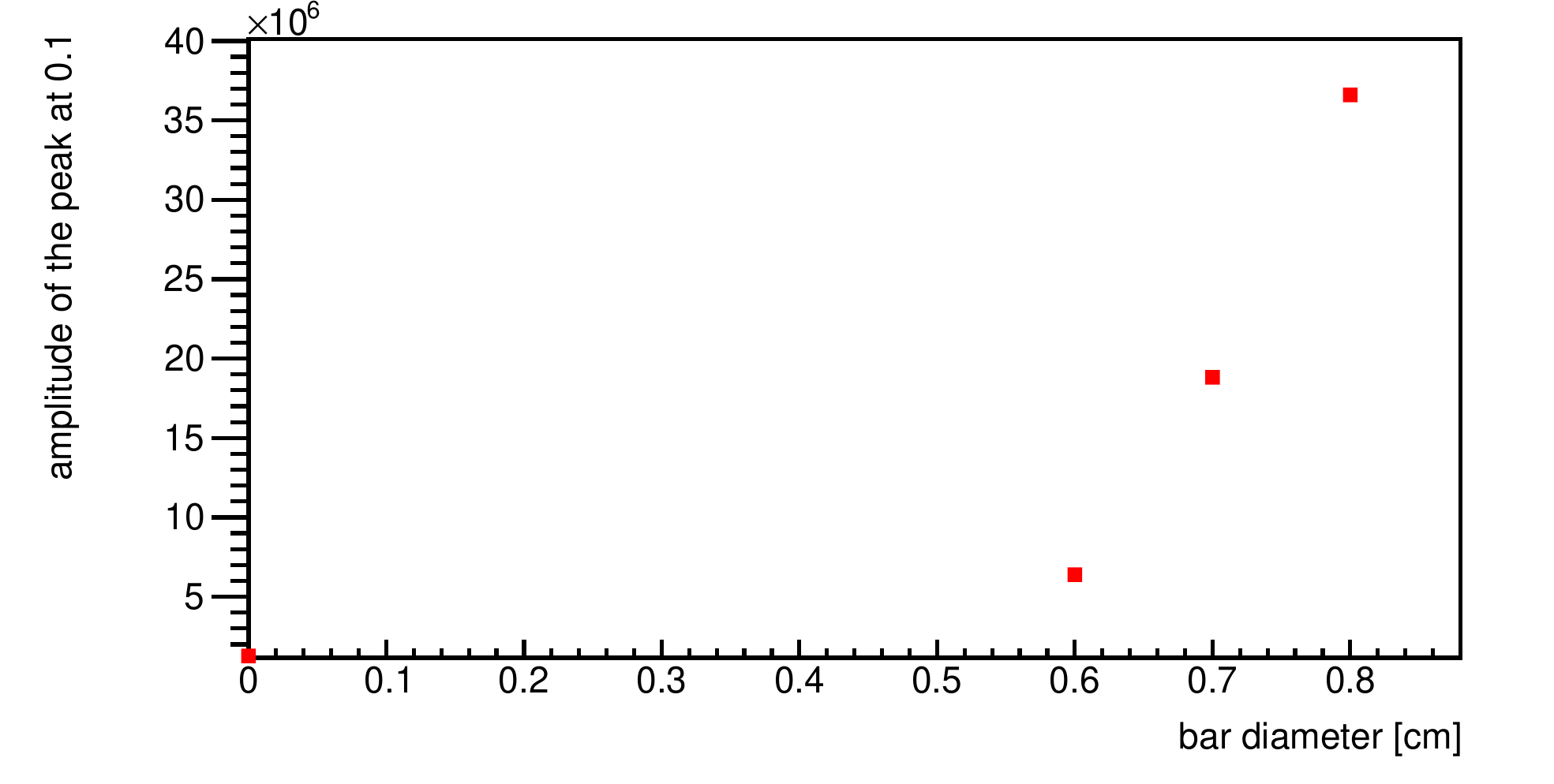}
	\caption{Amplitude of the peak at 0.1 of the normalized frequency distribution for different size of the bar diameter and spacing fixed at 10 cm. Amplitude of the bar with smallest commonly used diameters is almost 6 times higher than a background sample.}
	\label{fig:amplitude_0_1}
\end{figure}

The smallest bars in normal use have a 6~mm diameter.
In order to estimate whether the method is capable of finding such a small bar, the Fourier transform of grids with 8, 7 and 6~mm 
diameter bars were calculated using a 10~cm spacing and a one week worth of data taking. The results are shown in figure~\ref{fig: N2_1layer_mesh10_front_view_402M_30MT_frequency_0_6_0_7_0_8} and a zoomed version in figure~\ref{fig: N2_1layer_mesh10_front_view_402M_30MT_frequency_0_6_0_7_0_8_zoomed}.
Bars with 7 and 8~mm diameter at normalized frequency values of 0.1 and 0.9 are clearly distinguishable from the background. 
However, the signal for the 6~mm diameter case is less clear. 
Please note that the peak locations are determined by the spacing
and thus only peaks at the right location need to be considered.
To strengthen the 6~mm diameter signal, two weeks worth data taking were used. 
The peaks become clearer after two weeks of data taking, see 
figure~\ref{fig: N2_1layer_mesh10_front_view_804M_30MT_frequency_0_6_0_7_0_8} and a zoomed version in figure~\ref{fig: N2_1layer_mesh10_front_view_804M_30MT_frequency_0_6_0_7_0_8_zoomed}. Figure \ref{fig:amplitude_0_1} shows amplitude of the peak at 0.1 of the normalized frequency spectrum. 
Clearly, the method can detect the smallest size bar in use in practice for a 10~cm spacing. Amplitude of the bar with 6~mm diameter is almost 6 times higher than a background sample.\\

\begin{figure}[H]
	\centering
	\includegraphics[width=.95\columnwidth] {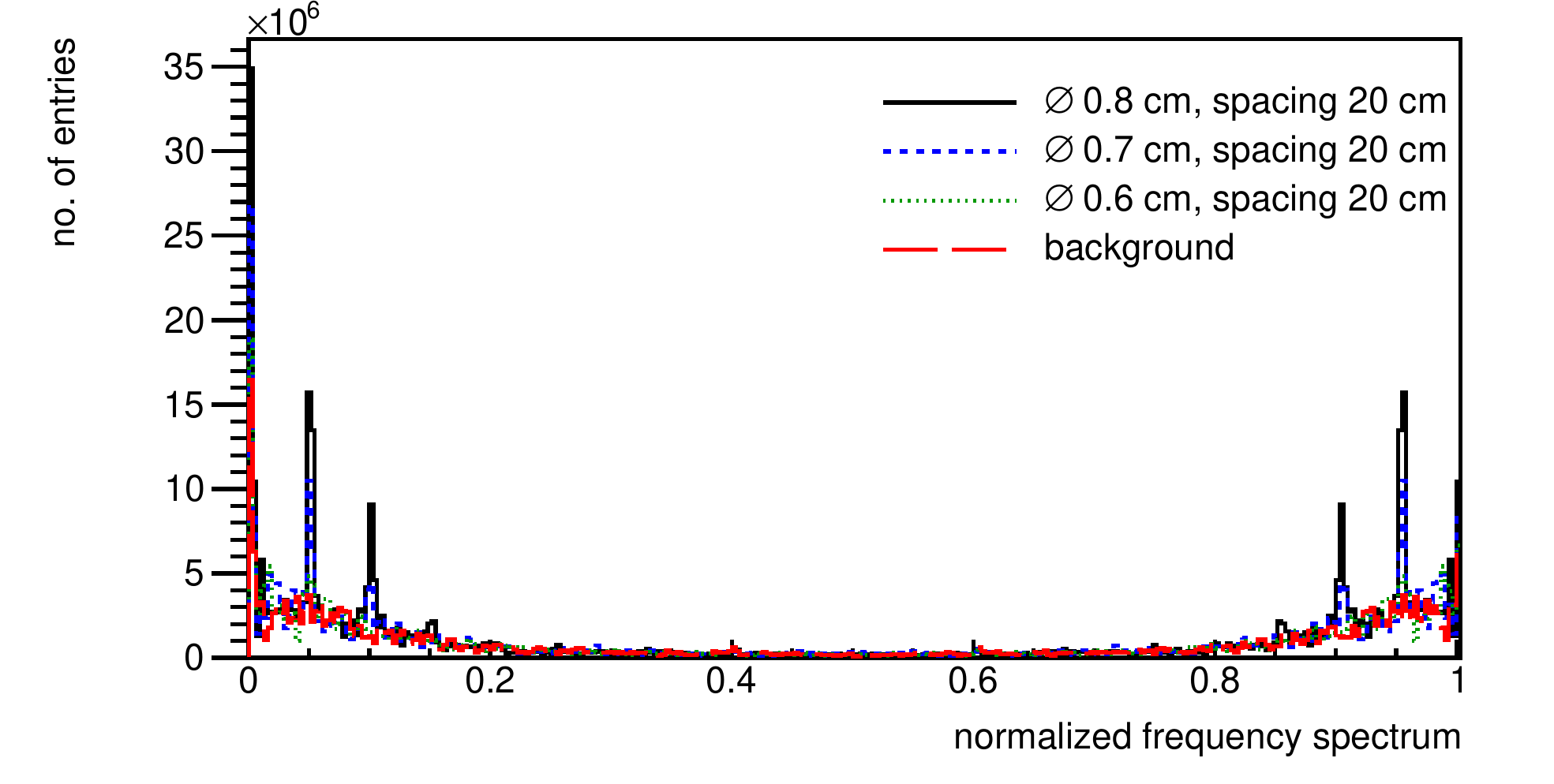}
	\caption{Normalized frequency spectrum for a reinforcement made of 6, 7 or 8~mm diameter bars and spacing of 20~cm.}
	\label{fig: N2_1layer_mesh20_front_view_402M_30MT_frequency_0_6_0_7_0_8}
\end{figure}

\begin{figure}[H]
	\centering
	\includegraphics[width=.95\columnwidth] {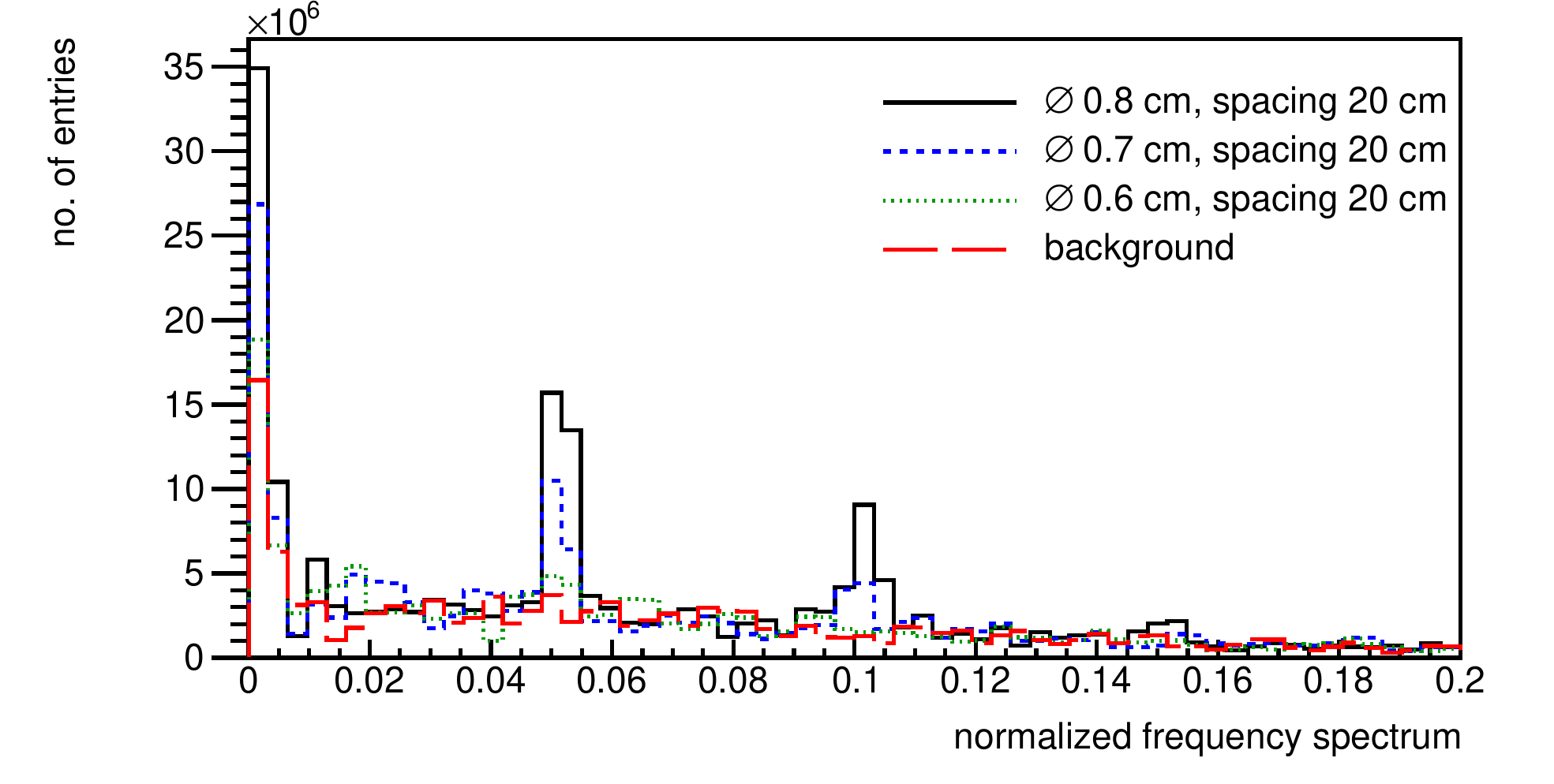}
	\caption{Zoomed, normalized frequency spectrum for a reinforcement made of 6, 7 or 8~mm diameter bars and spacing of 20~cm.}
	\label{fig: N2_1layer_mesh20_front_view_402M_30MT_frequency_0_6_0_7_0_8_zoomed}
\end{figure}

\begin{figure}[H]
	\centering
	\includegraphics[width=.95\columnwidth] {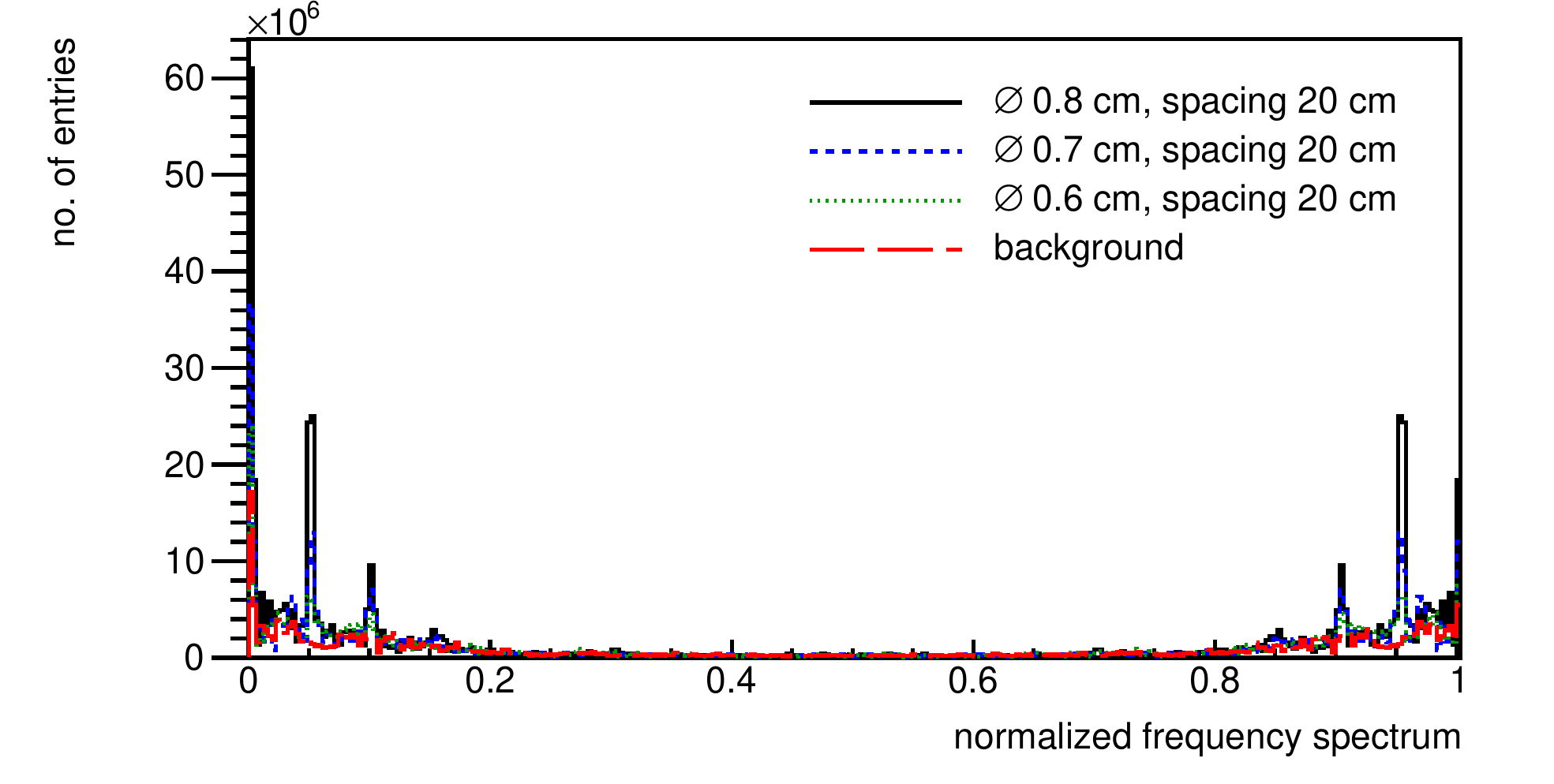}
	\caption{Normalized frequency spectrum for a reinforcement made of 6, 7 or 8~mm diameter bars and spacing of 20~cm. Time of data taking was increased to two weeks.}
	\label{fig: N2_1layer_mesh20_front_view_804M_30MT_frequency_0_6_0_7_0_8}
\end{figure}

\begin{figure}[H]
	\centering
	\includegraphics[width=.95\columnwidth] {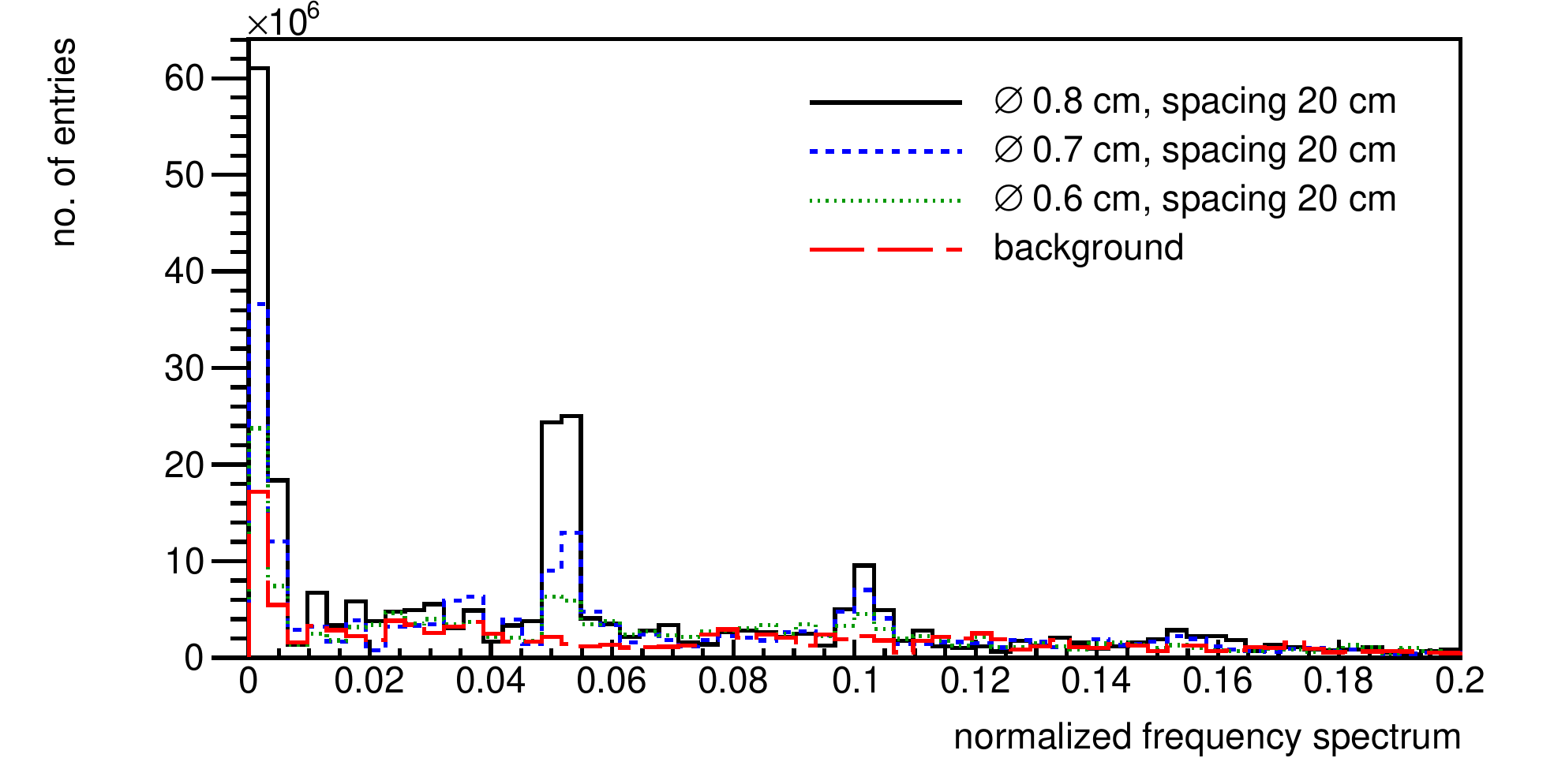}
	\caption{Zoomed, normalized frequency spectrum for a reinforcement made of 6, 7 or 8~mm diameter bars and spacing of 20~cm. Time of data taking was increased to two weeks.}
	\label{fig: N2_1layer_mesh20_front_view_804M_30MT_frequency_0_6_0_7_0_8_zoomed}
\end{figure}

To evaluate the limits of the method with respect to the spacing, a spacing of 20~cm was also considered. 
However, bigger spacing between bars results in a smaller amount of steel in the scanning area. This results in a smaller signal amplitude, as can be seen in figure~\ref{fig: N2_1layer_mesh20_front_view_402M_30MT_frequency_0_6_0_7_0_8}, zoomed version in figure~\ref{fig: N2_1layer_mesh20_front_view_402M_30MT_frequency_0_6_0_7_0_8_zoomed}.
The signal for 6~mm diameter bars is now almost below background level. 
However, when increasing the data taking time to two weeks (figure~\ref{fig: N2_1layer_mesh20_front_view_804M_30MT_frequency_0_6_0_7_0_8}, zoomed version in figure~\ref{fig: N2_1layer_mesh20_front_view_804M_30MT_frequency_0_6_0_7_0_8_zoomed}),
even the smallest bar is clearly visible again. \\

All of the scenarios presented in this publication used a single reinforcement grid. One possible scenario not shown here consists of multiple reinforcement grids. Adding more layers of grid will not reduce the ability of the detection in contrary to existing scanning methods. These geometries include more iron in the scanning area and thus the signal from steel is stronger and the time of data taking can be limited.

\section{Summary}
Inspection of ageing, reinforced concrete structures is a world-wide challenge and needs
novel non-destructive evaluation techniques with large penetration depths to precisely ascertain the
configuration of reinforcement and the internal condition of the structure and steelwork, which can possibly contain some impurities like voids.
Muon scattering tomography offers a technique that suits those needs. A method was presented to locate reinforcement placed in a large-scale concrete object.
The reinforcement was simulated as two layers of 2~m long bars, forming a grid, placed at a fixed distance from each other
inside a large concrete block. 
The technique exploits the periodicity of the bars in a reinforcement grid by 
considering the Fourier-transformed signal. The presence of a grid leads to peaks in the normalized Fourier frequency
spectrum. Peaks locations are determined by the grid spacing and their amplitude by the bar diameters. 
It is therefore possible to estimate both bar diameter and spacing with this method. 
Using only one week worth of data taking, bars with a diameter of  7~mm and larger, could easily be detected for a 
10~cm spacing. The signal for 6 mm diameter bar exceeds the background and but becomes very clear after two weeks of data taking. 
Increasing the spacing to 20~cm results in a smaller amount of iron in the scanning area, thus longer data taking is required.  
It has been shown that this method enables the detection of the smallest bars in practical use within one or two weeks of data
taking time and standard spacing. This is a very important result for non-destructive evaluation of civil structures.

\bibliographystyle{ieeetr}
\bibliography{bibfile}

\end{document}